\PassOptionsToPackage{unicode}{hyperref}
\PassOptionsToPackage{hyphens}{url}
\documentclass{article}
\usepackage{amsmath,amssymb}
\usepackage{iftex}
\ifPDFTeX
  \usepackage[T1]{fontenc}
  \usepackage[utf8]{inputenc}
  \usepackage{textcomp} % provide euro and other symbols
\else % if luatex or xetex
  \usepackage{unicode-math} % this also loads fontspec
  \defaultfontfeatures{Scale=MatchLowercase}
  \defaultfontfeatures[\rmfamily]{Ligatures=TeX,Scale=1}
\fi
\usepackage{lmodern}
\ifPDFTeX\else
\fi
\IfFileExists{upquote.sty}{\usepackage{upquote}}{}
\IfFileExists{microtype.sty}{% use microtype if available
  \usepackage[]{microtype}
  \UseMicrotypeSet[protrusion]{basicmath} % disable protrusion for tt fonts
}{}
\makeatletter
\@ifundefined{KOMAClassName}{% if non-KOMA class
  \IfFileExists{parskip.sty}{%
    \usepackage{parskip}
  }{% else
    \setlength{\parindent}{0pt}
    \setlength{\parskip}{6pt plus 2pt minus 1pt}}
}{% if KOMA class
  \KOMAoptions{parskip=half}}
\makeatother
\usepackage{xcolor}
\usepackage{graphicx}
\makeatletter
\def\maxwidth{\ifdim\Gin@nat@width>\linewidth\linewidth\else\Gin@nat@width\fi}
\def\maxheight{\ifdim\Gin@nat@height>\textheight\textheight\else\Gin@nat@height\fi}
\makeatother
\setkeys{Gin}{width=\maxwidth,height=\maxheight,keepaspectratio}
\makeatletter
\def\fps@figure{htbp}
\makeatother
\ifLuaTeX
  \usepackage{selnolig}  % disable illegal ligatures
\fi
\usepackage[margin=0.92in]{geometry}
\usepackage{booktabs,tabularx,threeparttable}
\usepackage{caption}
\usepackage{float}
\usepackage{fancyhdr}
\usepackage{setspace}
\usepackage[authoryear,round]{natbib}
\usepackage{bookmark}
\IfFileExists{xurl.sty}{\usepackage{xurl}}{} % add URL line breaks if available
\hypersetup{
  hidelinks,
  pdfcreator={LaTeX}}

\graphicspath{{./}}
\ifdefined\ANONYMOUS
  \hypersetup{pdfauthor={},pdfsubject={}}
\else
  \hypersetup{pdfauthor={Haojie Liu, Zihan Lin},pdfsubject={}}
\fi

\newcommand{\tabnote}[1]{\par\smallskip\parbox{.98\linewidth}{\footnotesize #1}}
\makeatletter
\def\@maketitle{%
  \newpage
  \null
  \vskip 2em%
  \begin{center}%
  \let\footnote\thanks
    {\Large\normalfont \@title \par}%
    \ifdefined\ANONYMOUS
      \vskip 1em%
    \else
      \vskip 1.5em%
      {\normalsize
        \lineskip .5em%
        \begin{tabular}[t]{c}%
          \@author
        \end{tabular}\par}%
      \vskip 1em%
    \fi
    {\normalsize \@date}%
  \end{center}%
  \par
  \vskip 1.5em}
\renewenvironment{abstract}{%
  \small
  \begin{center}%
    {Abstract\vspace{-.5em}\vspace{0pt}}%
  \end{center}%
  \quotation
}{\endquotation}
\makeatother

\renewcommand{\textbf}[1]{#1}

\usepackage{amsthm}
\newtheoremstyle{plainprop}{6pt}{6pt}{\itshape}{}{\normalfont}{.}{.5em}{}
\theoremstyle{plainprop}
\newtheorem{proposition}{Proposition}
\usepackage{enumitem}
\usepackage{flafter}
\newcommand{\NallFIGAROGainOne}{16.2}

\newcommand{\NallOECDGainOne}{5.0}

\newcommand{\Nautocorrelation}{0.10}
\newcommand{\NbothDropped}{BGR, DNK, HRV}
\newcommand{\NbothFIGAROGainFive}{11.1}
\newcommand{\NbothFIGAROGainOne}{16.2}
\newcommand{\NbothFIGAROLatestFive}{12.9}
\newcommand{\NbothFIGAROLatestOne}{5.5}

\newcommand{\NbothFIGAROPeersOne}{24.6}
\newcommand{\NbothList}{AUT, BEL, BGR, BRA, CAN, CYP, CZE, DEU, DNK, ESP, EST, FIN, FRA, GBR, GRC, HRV, HUN, IRL, ITA, LTU, LVA, NLD, POL, PRT, SVK, SVN, SWE, TUR, USA}
\newcommand{\NbothOECDGainFive}{8.5}
\newcommand{\NbothOECDGainOne}{13.0}
\newcommand{\NbothOECDLatestFive}{13.0}
\newcommand{\NbothOECDLatestOne}{6.8}

\newcommand{\NbothOECDPeersOne}{22.2}

\newcommand{\NchangeNoneNone}{3.8}
\newcommand{\NchangeNoneTable}{23.5}

\newcommand{\NchangeTableTable}{7.0}
\newcommand{\NcrossFOGainOne}{7.2}

\newcommand{\NcrossFOPositiveOne}{27}
\newcommand{\NcrossOFGainOne}{9.4}

\newcommand{\NcrossOFPositiveOne}{29}
\newcommand{\NcrossPanel}{14.7}

\newcommand{\NdecayRate}{15}
\newcommand{\NdeclineGain}{11.1}
\newcommand{\NdeclineGainHigh}{13.3}
\newcommand{\NdeclineGainLow}{8.9}
\newcommand{\NdistAllNineteen}{29.6}
\newcommand{\NdistAllTwentyTwo}{18.2}
\newcommand{\NdistNoTable}{11.9}
\newcommand{\NdistStayEighteen}{36.9}
\newcommand{\NdistStayNineteen}{39.2}
\newcommand{\NdistStayTwentyTwo}{36.8}
\newcommand{\NdistSwitchNineteen}{35.1}
\newcommand{\NdistSwitchTwentyTwo}{13.2}
\newcommand{\NdistTable}{25.4}

\newcommand{\Nelasticity}{0.52}
\newcommand{\NelasticityHigh}{0.57}
\newcommand{\NelasticityLow}{0.46}

\newcommand{\NeventIntensityDiff}{0.003}
\newcommand{\NeventIntensityHigh}{0.008}
\newcommand{\NeventIntensityLow}{\ensuremath{-}0.004}

\newcommand{\NeventKernelSwitch}{9.0}
\newcommand{\NeventRhoControl}{\ensuremath{-}0.006}
\newcommand{\NeventRhoDiff}{\ensuremath{-}0.055}
\newcommand{\NeventRhoHigh}{\ensuremath{-}0.013}
\newcommand{\NeventRhoLow}{\ensuremath{-}0.094}
\newcommand{\NeventRhoSwitch}{\ensuremath{-}0.061}
\newcommand{\NeventSupplierControl}{0.1}
\newcommand{\NeventSupplierDiff}{7.1}
\newcommand{\NeventSupplierHigh}{10.1}
\newcommand{\NeventSupplierLow}{3.7}
\newcommand{\NeventSupplierSwitch}{7.2}
\newcommand{\NfeRhoAll}{\ensuremath{-}0.020}
\newcommand{\NfeRhoAllHigh}{0.024}
\newcommand{\NfeRhoAllLow}{\ensuremath{-}0.067}
\newcommand{\NfeRhoClean}{\ensuremath{-}0.075}
\newcommand{\NfeRhoCleanHigh}{\ensuremath{-}0.017}
\newcommand{\NfeRhoCleanLow}{\ensuremath{-}0.135}
\newcommand{\NfeTable}{16.2}
\newcommand{\NfeTableHigh}{17.8}
\newcommand{\NfeTableLow}{14.5}
\newcommand{\NfigaroChangeDelivered}{6.0}
\newcommand{\NfigaroChangeEstimated}{1.5}
\newcommand{\NfigaroMedianDelivered}{4.3}
\newcommand{\NfigaroMedianEstimated}{0.5}
\newcommand{\NfigaroNearTemplate}{6}

\newcommand{\NgainOne}{14.5}
\newcommand{\NgainOneHigh}{16.4}
\newcommand{\NgainOneLow}{12.6}
\newcommand{\NgainOtherOne}{7.9}
\newcommand{\NgainOtherTen}{0.9}
\newcommand{\NgainTen}{3.4}
\newcommand{\NgainTenHigh}{5.5}
\newcommand{\NgainTenLow}{1.3}

\newcommand{\NhalfLife}{4.3}

\newcommand{\NhalfLifeHigh}{5.8}
\newcommand{\NhalfLifeLow}{3.0}
\newcommand{\NkazEighteen}{42}
\newcommand{\NkazNineteen}{10}
\newcommand{\NkazSeventeen}{9}
\newcommand{\NkernelAge}{6.5}

\newcommand{\NlatestFifteen}{24.1}
\newcommand{\NlatestFive}{14.7}

\newcommand{\NlatestOne}{6.6}
\newcommand{\NlatestOverOwnOne}{3.0}
\newcommand{\NlatestSixteen}{24.4}
\newcommand{\NlatestTen}{21.2}

\newcommand{\NlatestTwo}{9.2}
\newcommand{\NmFIGAROAllNineteen}{33.5}
\newcommand{\NmFIGAROAllTwentyTwo}{33.9}

\newcommand{\NmOECDAllNineteen}{32.4}
\newcommand{\NmOECDAllTwentyTwo}{15.4}
\newcommand{\NmOECDStayNineteen}{31.4}
\newcommand{\NmOECDStayTwentyTwo}{31.6}
\newcommand{\NmanufacturingShare}{24}

\newcommand{\NmultFIGARO}{1.563}
\newcommand{\NmultOECD}{1.564}

\newcommand{\NnBoth}{29}
\newcommand{\NnBothStrict}{26}
\newcommand{\NnCleanMult}{31}
\newcommand{\NnControlObs}{54}
\newcommand{\NnCountryYears}{2,240}
\newcommand{\NnFigaroEstimated}{60}

\newcommand{\NnLikeNineteen}{29}

\newcommand{\NnLikeTwentyTwo}{67}

\newcommand{\NnPanel}{25}
\newcommand{\NnPriceYears}{2,160}
\newcommand{\NnRest}{55}
\newcommand{\NnStay}{11}

\newcommand{\NnSwitch}{40}
\newcommand{\NnSwitchEvents}{34}

\newcommand{\NnTableAll}{1,237}
\newcommand{\NnTablesNineteen}{52}

\newcommand{\NnTablesTwentyTwo}{20}

\newcommand{\NnoiseShare}{51}
\newcommand{\NoffDiagonalShare}{83}
\newcommand{\NoldBuyerNineteen}{37.6}
\newcommand{\NoldBuyerTwentyTwo}{29.0}

\newcommand{\NoldGainOne}{7.8}
\newcommand{\NoldGainTen}{0.5}
\newcommand{\NoldPositiveTen}{44}
\newcommand{\NoldTransNineteen}{45.7}
\newcommand{\NoldTransTwentyTwo}{41.6}

\newcommand{\NownOne}{9.6}

\newcommand{\NpanelDropped}{DNK}
\newcommand{\NpanelList}{AUT, BEL, BRA, CAN, CMR, COL, CYP, CZE, DEU, DNK, ESP, EST, FIN, FRA, GBR, HUN, IRL, ITA, JPN, LTU, LUX, NLD, SEN, SVK, USA}

\newcommand{\NpeersOne}{24.1}

\newcommand{\NpeersTen}{25.0}

\newcommand{\NpooledOne}{24.7}

\newcommand{\NpositiveTen}{20}

\newcommand{\NpriceKernel}{0.04}
\newcommand{\NpriceTransactions}{2.7}

\newcommand{\NrasLatestFive}{10.6}
\newcommand{\NrasLatestOne}{4.6}
\newcommand{\NrasLatestTen}{15.7}
\newcommand{\NrasPeers}{19.3}
\newcommand{\NrasTemplate}{19.6}
\newcommand{\NratioArrive}{3.4}
\newcommand{\NratioOne}{3.7}

\newcommand{\NrhoCrossPanel}{0.82}
\newcommand{\NrhoEarlyMax}{0.81}
\newcommand{\NrhoEarlyMin}{0.78}
\newcommand{\NrhoLagEight}{0.82}

\newcommand{\NrhoTwentyOne}{0.74}
\newcommand{\NrhoTwentyTwo}{0.72}
\newcommand{\NrobustHighOne}{14.7}
\newcommand{\NrobustHighTen}{3.4}
\newcommand{\NrobustLowOne}{11.6}
\newcommand{\NrobustLowTen}{1.9}
\newcommand{\NshareLikeNoTable}{88}
\newcommand{\NshareLikeTable}{3.6}
\newcommand{\NshareLikeTableTwentyOne}{38}
\newcommand{\NshareLikeTableTwentyTwo}{65}

\newcommand{\NshrinkGainEight}{0.8}

\newcommand{\NshrinkGainTen}{1.2}
\newcommand{\NshrinkGainTenHigh}{1.8}
\newcommand{\NshrinkGainTenLow}{0.6}

\newcommand{\NstayList}{BRN, CMR, COD, CZE, GBR, JPN, NLD, PER, SEN, STP, TWN}

\newcommand{\Ntemplate}{24.4}
\newcommand{\NtemplateDrift}{1.9}
\newcommand{\NtemplateErrorBound}{16.6}
\newcommand{\NtemplateMinusLatestTen}{3.1}
\newcommand{\NtemplateMinusLatestTenHigh}{5.5}
\newcommand{\NtemplateMinusLatestTenLow}{0.7}
\newcommand{\NtemplateMinusLatestTenPositive}{19}
\newcommand{\NtemplatePairNineteen}{13.9}

\newcommand{\NtemplatePanelMean}{24.2}

\newcommand{\NtransportShare}{19}
\newcommand{\NusAusNineteen}{33}
\newcommand{\NusAusTwentyTwo}{8}
\newcommand{\NusChangeTwentyTwo}{25}

\newcommand{\NusFinNineteen}{31}
\newcommand{\NusFinTwentyTwo}{7}
\newcommand{\NusOwnFigaro}{0.03}

\newcommand{\NusOwnTwentyTwo}{15.0}

\newcommand{\NvarCountryMeasured}{66}

\newcommand{\NnetDeAgFood}{52.8}
\newcommand{\NnetJpAgFood}{38.1}
\newcommand{\NnetDeMinerals}{24.2}
\newcommand{\NnetJpMinerals}{23.0}
\newcommand{\NnetDeAir}{8.1}
\newcommand{\NnetJpAir}{55.4}
\newcommand{\NnetDeWater}{30.0}
\newcommand{\NnetJpWater}{9.2}
\newcommand{\NnetDeEstate}{35.2}
\newcommand{\NnetJpEstate}{18.3}

\title{Persistence and Change in National Production Networks}
\ifdefined\ANONYMOUS
  \author{}
\else
  \author{Haojie Liu\\ University of California, Riverside\\ hliu332@ucr.edu \and Zihan Lin\\ University of California, Riverside\\ zlin169@ucr.edu}
\fi
\date{19 November 2026}
\hypersetup{pdftitle={Persistence and Change in National Production Networks}}
\begin{document}
\maketitle

\begin{abstract}
Many country-years in multi-country input-output databases have no national table. We study the kernel of domestic intermediate transactions, the part that rescaling to new industry totals cannot update. Among \NnPanel{} economies with a national table every year, the latest table misplaces \NlatestOne{} percent of the kernel one year ahead and \NlatestTen{} percent ten years ahead, while the five most similar economies misplace \NpeersOne{} to \NpeersTen{} percent at every horizon. In the OECD tables, \NshareLikeNoTable{} percent of country-years without a national table lie close to one cross-country template, against \NshareLikeTable{} percent of country-years with a table. The template spreads from \NnLikeNineteen{} economies in 2019 to \NnLikeTwentyTwo{} in 2022. In this sample an own table's error meets its distance to that year's template near fifteen years, not as a general shelf life.
\end{abstract}


\section{Introduction}\label{sec:intro}
Input-output tables are the empirical basis of most quantitative work on production networks. Multisector models are calibrated to them \citep{acemoglu2012,caliendoparro,baqaeefarhi}, measures of global value chains are computed from them \citep{johnson2018}, and multipliers, footprints and exposure indicators inherit their structure. The tables are also among the slowest statistics to produce. National supply and use tables often appear three years or more after the reference year \citep{oecd2026}, and many statistical offices compile them only for benchmark years. Two practices follow. Researchers use the most recent table to describe later years, and the compilers of multi-country databases estimate tables for country-years in which no national table exists. Each practice rests on an assumption about persistence that has rarely been tested: that an economy's own past structure remains a good guide, or that the structure of other economies is a good substitute. This paper measures how good each guide is and documents what a widely used database contains in the years that had to be filled.

A single cell shows why the second issue matters. In the 2025 edition of the OECD Inter-Country Input-Output tables, the US construction industry buys less than 0.1 percent of its domestic intermediate inputs from itself in every year from 1997 to 2021. In 2022 the share is \NusOwnTwentyTwo{} percent. Eurostat's FIGARO tables, which obtain US supply and use tables through the OECD, show \NusOwnFigaro{} percent for the same year. The jump is not confined to one cell or one country. Once industry sizes are removed, the 2022 table for the United States differs from its own 2021 table by \NusChangeTwentyTwo{} percent of its mass, five times the usual annual change. It lies within \NusFinTwentyTwo{} percent of the table for Finland and \NusAusTwentyTwo{} percent of the table for Australia, economies from which it had differed by \NusFinNineteen{} and \NusAusNineteen{} percent in 2019. The OECD's documentation lists no national table for the United States in 2022.

To study structure apart from size we work with the kernel of each table: the matrix obtained by scaling domestic intermediate transactions so that every industry has the same total sales and the same total purchases. The kernel has two properties that make it the natural object for both questions. First, it is exactly the part of a table that biproportional updating, the RAS method routinely used to carry a table to a new year, leaves unchanged. The kernel distance measures how the sourcing pattern differs once row and column scale factors have been removed. When the target industry totals are uniform, that distance equals the total variation error of updating the old table to those totals by RAS. With the uneven totals of an actual year the two quantities differ, and Section~\ref{sec:persistence} reports the second one separately. Second, when buyers have CES technologies with a common elasticity of substitution, prices and industry sizes enter transactions as row and column factors, so the kernel depends only on the input weights of the technology. Movements in the kernel are changes in who supplies whom that prices and scale do not account for. We measure the distance between kernels by total variation and report it in percent: a distance of 10 means that one tenth of the kernel's mass sits in different cells.

The evidence comes from a forecasting experiment and from the documentation of two databases. For each economy and each horizon from one to ten years we forecast the kernel from information dated at the forecast origin: the economy's latest table, the mean of its last five tables, and the average of the five most similar other economies over the same five years. Comparing own and foreign histories at equal dates shows how much country-specific information a table carries and how fast that information depreciates. We run the experiment on the \NnPanel{} economies for which the OECD lists a national supply and use or input-output table in every year from 2000 to 2020, on the remaining \NnRest{} economies, and on a matched sample of 42 economies and 33 industries that is also available in FIGARO.

Where national tables exist every year, structure is strongly country specific and drifts slowly. One year ahead, the latest table misplaces \NlatestOne{} percent of the kernel and the five nearest peers misplace \NpeersOne{} percent. The peers' error is almost the same at every horizon, whereas the own error grows roughly with the square root of the table's age and reaches \NlatestTen{} percent at ten years. The advantage of an economy's own five-year history over its peers falls from \NgainOne{} to \NgainTen{} percentage points between one and ten years, a half-life of about four years, and it is still positive in \NpositiveTen{} of \NnPanel{} economies at the longest horizon. The latest table beats the five-year mean at every horizon, so the drift shows no tendency to revert. The OECD and FIGARO tables give similar profiles for the \NnBoth{} economies with national tables in both, and a history compiled by one agency forecasts the other agency's table for the same economy better than foreign histories do. Country-specific structure therefore belongs to the economies themselves and survives a change of compiler.

Country-years without a national table tell a different story. In the OECD tables, \NshareLikeNoTable{} percent of them have kernels within 16 percent of a single template, the average kernel of that year's unmeasured economies. Among country-years with a national table the share is \NshareLikeTable{} percent. Economies switch between the two states as their tables arrive and lapse. Kazakhstan, which has a national table for 2018 and none for 2017 or 2019, sits \NkazSeventeen{}, \NkazEighteen{} and \NkazNineteen{} percent from the template in those three years. Coverage thins toward the end of the release. The OECD lists national tables for \NnTablesNineteen{} of 80 economies in 2019 and for \NnTablesTwentyTwo{} in 2022, and the number of template-like economies rises from \NnLikeNineteen{} to \NnLikeTwentyTwo{}. Eurostat fills the same kind of gap differently. Its estimated tables for non-EU partners carry forward the economy's own structure, so the kernel changes by \NfigaroMedianEstimated{} percent in a typical estimated year, against \NfigaroMedianDelivered{} percent for tables delivered by national statistical institutes.

These filling rules shape measured change. In the OECD tables the average distance between the kernels of the 80 economies falls from \NdistAllNineteen{} to \NdistAllTwentyTwo{} percent between 2019 and 2022. Among the \NnSwitch{} economies that become template-like it falls from \NdistSwitchNineteen{} to \NdistSwitchTwentyTwo{} percent, among the \NnStay{} that stay specific it moves from \NdistStayNineteen{} to \NdistStayTwentyTwo{} percent, and in FIGARO it does not move. Forecast evaluations that pool both kinds of country-year understate the own-history advantage by almost half and can make foreign histories look as informative as an economy's own. In a standard application, agreement between the two databases on the ranking of sectors by domestic output multiplier is stable at about 0.8 from 2010 to 2020 and drops to \NrhoTwentyTwo{} in 2022. In the year an economy's OECD table becomes template-like, the two databases diverge on who supplies whom while continuing to agree on how much each industry buys.

The experiment also ranks the two filling rules. Measured against a national table, the template is off by \Ntemplate{} percent in the year it is used. A national table carried forward is off by \NlatestFive{} percent after five years and \NlatestTen{} percent after ten. In this sample an own table carried forward has the smaller kernel error on the short side of a crossing near fifteen years. Horizons past ten years rest on few origins, and the template uses other economies' kernels from the target year, so the crossing is a description of this period rather than a shelf life or an updating rule. For users, the practical step is to check which country-years the compilers list as resting on national tables before interpreting changes in structure, and to treat the last two reference years of a release as provisional. Distance from the template is a check on that list. The template is initialized from the country-years the OECD records as having no national table, so the check is not independent of the documentation.

\section{Related Literature}\label{sec:lit}
The paper connects three literatures. The first studies how production networks form and change. \citet{carvalhovoigt} show that existing input links predict which new inputs an industry adopts, \citet{oberfield} and \citet{acemogluazar} model the choice of suppliers and inputs, and \citet{mundt} estimates link formation among European industries. Firm-level data reveal how often individual supplier relations are rewired \citep{reisch2026} and how much is lost when firms are aggregated into industries \citep{diem}. At the level of national tables, \citet{chenery1958} established that economies share a common core of interindustry structure, \citet{nordlund} and \citet{haddad} compare national structures and their trajectories, and \citet{barros2026} relate changes in input composition to structural transformation. Quantitative models usually fix the network at a base-year table, which amounts to assuming constant expenditure shares, and \citet{atalay} and \citet{baqaeefarhi} show that departures from that assumption matter for aggregate outcomes. We add a measure of how much country-specific structure a table contains and how long it lasts, on a common footing for many economies and two compilers.

The second literature concerns updating. Biproportional adjustment goes back to \citet{stone1961} and \citet{bacharach1970}. \citet{lahr2004} survey its variants, and \citet{temurshoev2011} test eight updating methods against realized supply and use tables for the Netherlands and Spain. Scenario-based projections \citep{beaufils2022} and near-real-time tables \citep{huo2022} extend the idea to complete multi-regional systems, and official practice is described in \citet{unhandbook}. All of these methods take an interaction structure from somewhere, an older table of the same economy or a table of another economy, and fit it to current totals. We evaluate that choice directly. Proposition~\ref{prop:kernel} shows that RAS recovers the new table exactly when the kernels coincide, and that the kernel distance equals the remaining total variation error when the target totals are uniform. The experiment measures that distance by horizon and by source of the structure. The design follows rolling-origin practice \citep{tashman}, the combinations we report are fixed-weight averages \citep{timmermann2006}, and \citet{pesaran2026} analyze the trade-off between pooling and heterogeneity that the comparison of own and foreign histories embodies.

The third literature compares multi-regional databases. \citet{tarne}, \citet{owen2014} and \citet{pinero} document differences between databases in transaction structure, footprints and trade indicators, \citet{lenzen2011} studies the role of aggregation, \citet{wiebe2016} find that assumptions about missing or conflicting data matter more for multi-regional results than whether the global matrices are balanced by RAS, and \citet{badr2026} review the sources of uncertainty. \citet{johnson2013} show for the Penn World Table that revisions across vintages change growth estimates. We add a specific and checkable source of disagreement. Using the compilers' own documentation of which country-years rest on national tables \citep{oecdnotes2025,yamano,eurostatmeta}, we show that unmeasured country-years in one database share a common kernel, that this accounts for a prominent change in the most recent years, and that the two databases give similar results where both are measured.

\section{Data and Measurement}\label{sec:data}
\subsection{Sources}
The main source is the 2025 edition of the OECD Inter-Country Input-Output tables \citep{oecd2025}, which covers 80 economies and 50 industries from 1995 to 2022. We extract each economy's domestic block of intermediate transactions at current prices. Scaling to equal totals requires every industry to buy and to sell something, so we keep the 34 industries with positive domestic purchases and sales in every economy and year. They account for 75.8 percent of domestic intermediate transactions on average. The industries dropped include mining, refining, machinery and motor vehicles, and our results do not speak to them (Appendix~\ref{app:data}). The second source is Eurostat's FIGARO tables, 2026 edition \citep{eurostat}. For the comparison of databases we aggregate both sources to 33 common industries for the 42 economies they share, from 2010 to 2022.

The documentation of both databases records which country-years had a national source available. For the OECD, the supplementary notes to the 2025 edition \citep{oecdnotes2025} list for every economy and year whether a national supply and use table, an input-output table, both, or neither was available. The list is about availability to the compiler. It does not say which industry cells were taken from that source. Balancing and interpolation can still change cells in a year coded as having a table. Of the \NnCountryYears{} country-years, \NnTableAll{} have at least one national table. For the others the methodology describes two procedures \citep{yamano}. Where use tables are not published for a year, ``the structure of input coefficients of the intermediate section are estimated using interpolated numbers from the nearest available years' structures''. Where ``sectoral details are not available for a specific country, average production structures of other countries are used as initial values to initiate the balancing procedure''. Which procedure applies to which country-year is not reported. For FIGARO, the metadata list the origin of the supply and use tables of each non-EU partner by year: a national statistical institute, the OECD, or tables ``estimated by Eurostat using SUTs from other years'' \citep{eurostatmeta}. The published tables are also harmonized and balanced. Tables for EU member states, Norway and T\"urkiye are delivered by national institutes under the European transmission program.

\subsection{The kernel}
Let $Z_{ct}$ be the $m\times m$ matrix of domestic intermediate transactions of economy $c$ in year $t$, with suppliers in rows and buyers in columns. Its kernel is
\begin{equation}
Q_{ct}=D^r_{ct}\,Z_{ct}\,D^c_{ct},\qquad Q_{ct}\,\iota=Q_{ct}'\,\iota=\tfrac1m\,\iota,
\label{eq:q}
\end{equation}
where $D^r_{ct}$ and $D^c_{ct}$ are diagonal matrices with positive diagonals and $\iota$ is a vector of ones. The kernel exists and is unique for every matrix in our samples \citep{sinkhorn,chakrabarty}. It has total mass one and keeps the zeros of $Z_{ct}$. We compare kernels by total variation,
\begin{equation}
d(A,B)=\tfrac12\sum_{i,j}|A_{ij}-B_{ij}|,
\label{eq:tv}
\end{equation}
the share of mass that must be moved to turn $A$ into $B$, and we report $100\,d$.

\begin{proposition}\label{prop:kernel}
Let $Z$ and $\tilde Z$ be nonnegative matrices with the same zero pattern for which kernels exist, and write $Q(\cdot)$ for the kernel.
\begin{enumerate}[label=(\alph*),leftmargin=2em,itemsep=1pt,topsep=2pt]
\item Invariance. $Q(RZS)=Q(Z)$ for all diagonal matrices $R$ and $S$ with positive diagonals.
\item Updating. The biproportional adjustment of $Z$ to the row and column totals of $\tilde Z$ equals $\tilde Z$ if and only if $Q(Z)=Q(\tilde Z)$. If the totals of $\tilde Z$ are uniform, the total variation error of the adjustment is $d(Q(Z),Q(\tilde Z))$.
\item Technology. Suppose each buyer $j$ minimizes the cost of a CES bundle of intermediate inputs with weights $a_{ij}$ and an elasticity of substitution $\sigma$ common to all buyers, at supplier prices $p_i$. Then $Z=\operatorname{diag}(p_i^{1-\sigma})\,A\,\operatorname{diag}(P_j^{\sigma}M_j)$, where $A=(a_{ij})$ and $P_j$ and $M_j$ are the price and quantity of buyer $j$'s bundle. Hence $Q(Z)=Q(A)$.
\end{enumerate}
\end{proposition}

Appendix~\ref{app:proof} gives the proof. Part (a) says that the kernel ignores anything that acts on a whole row or a whole column: the size of an industry as a seller or as a buyer, a price change common to all buyers of a product, the share of intermediate inputs in a buyer's costs, or import penetration that is uniform across the buyers of a product. What it keeps are the cross-product ratios $Z_{ij}Z_{kl}/(Z_{il}Z_{kj})$, the statistics that separate bilateral frictions from exporter and importer terms in gravity models \citep{headries} and that have long been used to compare association across contingency tables \citep{mosteller}.

Part (b) links the kernel to practice. RAS finds row and column multipliers that fit an old table to new totals. By (b) it recovers the new table exactly when the two kernels coincide. The kernel distance equals the total variation error of that update when the new totals are uniform. In general it measures the structural difference that remains after row and column factors are removed, which need not equal the total variation error of the RAS-updated transaction matrix. \citet{stone1961} read the row and column multipliers as substitution and fabrication effects. In that language the kernel distance measures structural change beyond uniform substitution and fabrication.

Part (c) gives the economic content. In the CES models commonly used for production networks \citep[for example][]{atalay,baqaeefarhi}, the kernel is a function of the technology weights alone, whatever the prices and the sizes of industries. With buyer-specific elasticities $\sigma_j$ the statement holds up to an interaction between $\sigma_j-\bar\sigma$ and relative supplier prices. The data bear out the invariance to prices. The OECD publishes each table at current prices and at the prices of the previous year. Across \NnPriceYears{} country-years the two versions differ by \NpriceTransactions{} percent in transaction shares and by \NpriceKernel{} percent in the kernel.

\subsection{Forecasts}
A forecast origin is a year $\tau$ and the target is $\tau+h$. We consider three forecasts of $Q_{c,\tau+h}$ that use only tables dated $\tau$ or earlier. The latest own table is $Q_{c\tau}$. The own five-year mean is $B_{c\tau}=\frac15\sum_{s=0}^{4}Q_{c,\tau-s}$. The peer forecast $F_{c\tau}$ averages $B_{k\tau}$ over the five other economies $k$ whose $B_{k\tau}$ is closest to $B_{c\tau}$ in Frobenius distance, with weights proportional to inverse squared distance. Peers and weights are fixed at $\tau$, so own and peer histories end in the same year and have the same length. The own-history premium is
\begin{equation}
G_{c\tau h}=d(Q_{c,\tau+h},F_{c\tau})-d(Q_{c,\tau+h},B_{c\tau}),
\label{eq:g}
\end{equation}
which is positive when the economy's own history forecasts better.

We report horizons of 1, 2, 3, 5, 8 and 10 years and hold the target years fixed at 2014 to 2020 for every horizon. Horizons then differ in the dating of the history and not in the years being forecast. The identity $t=\tau+h$ rules out separating origin, target and horizon effects without a restriction, and fixing the targets is the restriction we impose. All tables come from one release of each database, so the exercise is retrospective. It asks how well earlier reference years of the current release predict later ones.

Intervals have nominal 95 percent coverage. Unless noted they come from two-way clustering by economy and target year \citep{cameron,mackinnon}. A bootstrap that resamples whole economies and a joint bootstrap over economies and blocks of adjacent years give similar answers (Appendix~\ref{app:forecast}).

\subsection{Measured economies}
Section~\ref{sec:template} shows that country-years with and without national tables differ in kind. We therefore estimate persistence on economies that are measured throughout: the \NnPanel{} economies for which the OECD lists a national table in every year from 2000 to 2020. Peers are drawn from the same group. The definition uses only the published documentation. The panel consists of 18 European economies together with Brazil, Cameroon, Canada, Colombia, Japan, Senegal and the United States. For the comparison of databases we use the \NnBoth{} economies that have an OECD-listed national table in every year from 2010 to 2020 and whose FIGARO tables are never Eurostat estimates over the same years. Appendix~\ref{app:data} lists both groups.

\section{Persistence Where National Tables Exist}\label{sec:persistence}
\subsection{The horizon profile}
Table~\ref{tab:horizon} reports the experiment for the \NnPanel{} measured economies. One year ahead the latest table misplaces \NlatestOne{} percent of the kernel and the five-year mean \NownOne{} percent. The five nearest peers misplace \NpeersOne{} percent, \NratioOne{} times the error of the latest table. The premium of own over foreign history is \NgainOne{} points, with an interval of [\NgainOneLow, \NgainOneHigh], and it is positive in every economy.

\begin{table}[!htbp]\centering\small
\caption{Forecast losses for economies with a national table every year}\label{tab:horizon}
\setlength{\tabcolsep}{5pt}
\begin{tabular}{rrrrrrlr}\toprule
 & \multicolumn{4}{c}{Loss of the forecast} & \multicolumn{3}{c}{Own five-year mean against peers}\\
\cmidrule(lr){2-5}\cmidrule(lr){6-8}
Years ahead & Own latest & Own five-year & Five peers & Template & Premium & 95\% interval & Positive \\
\midrule
1 & 6.6 & 9.6 & 24.1 & 24.4 & 14.5 & [12.6, 16.4] & 25/25 \\
2 & 9.2 & 11.5 & 24.1 & 24.4 & 12.7 & [10.8, 14.5] & 25/25 \\
3 & 11.1 & 13.1 & 24.2 & 24.4 & 11.1 & [9.2, 13.0] & 25/25 \\
5 & 14.7 & 16.1 & 24.4 & 24.4 & 8.3 & [6.3, 10.3] & 24/25 \\
8 & 19.1 & 19.8 & 24.7 & 24.4 & 5.0 & [2.8, 7.2] & 23/25 \\
10 & 21.2 & 21.6 & 25.0 & 24.4 & 3.4 & [1.3, 5.5] & 20/25 \\

\bottomrule\end{tabular}
\tabnote{\NnPanel{} economies, 34 industries, target years 2014 to 2020. Losses are total variation distances from the target kernel in percent. Own and peer forecasts use tables dated at or before the origin, which is the target year minus the horizon. The template is the kernel defined in Section~\ref{sec:template} for the target year. The premium is the peer loss minus the loss of the own five-year mean, with a two-way clustered interval. The last column counts economies with a positive mean premium.}
\end{table}

Two features of the table organize the rest of the paper. The first is that the peer error barely depends on the horizon. It is \NpeersOne{} percent at one year and \NpeersTen{} percent at ten. What an economy shares with similar economies is a common structure that does not age. The second is that the own error grows steadily, to \NlatestFive{} percent at five years and \NlatestTen{} percent at ten, because what is specific to an economy drifts. The premium therefore declines by \NdeclineGain{} points between one and ten years [\NdeclineGainLow, \NdeclineGainHigh]. A geometric fit to the ten annual horizons gives a decline of \NdecayRate{} percent per year, a half-life of \NhalfLife{} years [\NhalfLifeLow, \NhalfLifeHigh]. At ten years the premium is \NgainTen{} points [\NgainTenLow, \NgainTenHigh] and positive in \NpositiveTen{} of \NnPanel{} economies.

\begin{figure}[!htbp]\centering
\includegraphics[width=\linewidth]{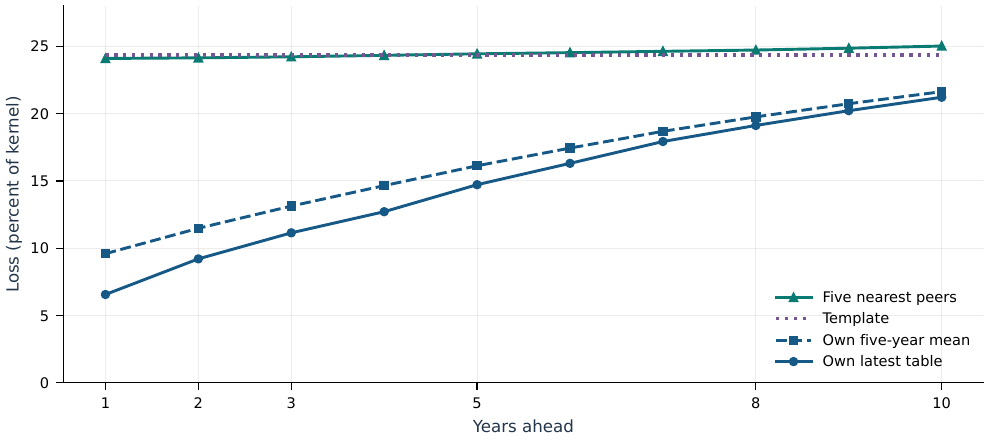}
\caption{Forecast losses by horizon for the \NnPanel{} economies with a national table every year. The losses are those in Table~\ref{tab:horizon}, for every horizon from one to ten years.}
\label{fig:horizon}
\end{figure}

The latest table is the better own forecast at every horizon. Averaging five years would help if year-to-year movements were noise around a stable national structure. It costs \NlatestOverOwnOne{} points at one year, and the cost shrinks to almost nothing at ten.

\subsection{Drift}
\begin{figure}[!htbp]\centering
\includegraphics[width=\linewidth]{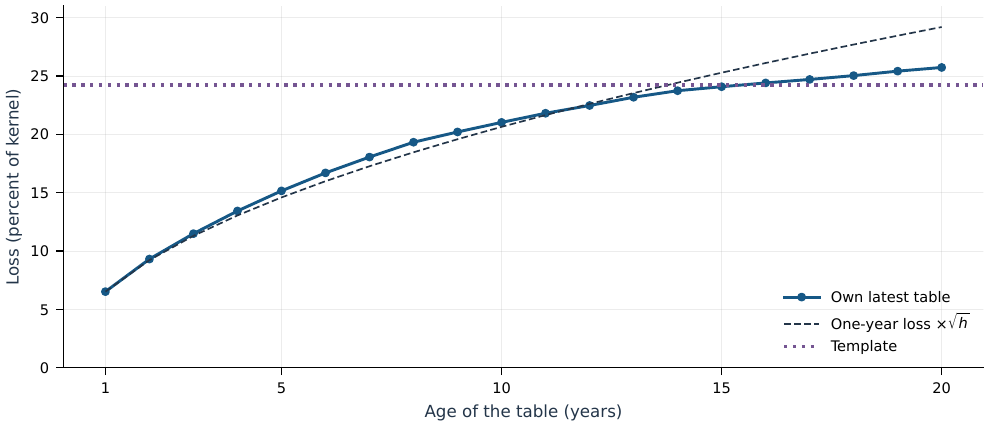}
\caption{Loss of the latest own table as it ages, for the \NnPanel{} economies and every pair of years between 2000 and 2020. The dashed curve multiplies the one-year loss by the square root of the age, and the dotted line is the mean distance of these economies' kernels from the template.}
\label{fig:long}
\end{figure}

Figure~\ref{fig:long} follows the latest table to an age of twenty years. Over the first ten years the loss rises with an elasticity of \Nelasticity{} with respect to age [\NelasticityLow, \NelasticityHigh], close to the square-root growth expected when annual changes accumulate without a systematic direction. Annual changes are slightly negatively correlated from one year to the next ($-$\Nautocorrelation{}), and a linear fit of squared distance on age attributes about half of the one-year change to a transitory component (Appendix~\ref{app:forecast}). A persistent component with independent increments plus some transitory noise describes the data well. Over horizons of up to ten years there is no sign of a fixed national structure to which a table returns, and the most recent table is the right one to carry forward.

The curve reaches the template line at about fifteen years. The loss is \NlatestFifteen{} percent at fifteen years and \NlatestSixteen{} percent at sixteen, against a mean distance of \NtemplatePanelMean{} percent between these economies' kernels and the template. Horizons beyond ten years rest on few origin years, so the crossing point is approximate. The template at each age is the template of the target year, built from other economies' kernels in that year, so the two sides of the comparison do not use the same information. The crossing describes the \NnPanel{} economies between 2000 and 2020. It is not a general date at which a table stops being informative, and it is not a rule for real-time updating.

For quantitative work the profile gives a rule of thumb. A model calibrated to a table that is $h$ years old carries a kernel error of about \NlatestOne{}$\sqrt h$ percent: \NlatestTwo{} percent after two years, \NlatestFive{} after five and \NlatestTen{} after ten.

\subsection{What is specific to an economy}
No foreign table is a good substitute for an economy's own. The five nearest peers miss the one-year target by \NpeersOne{} percent, the unweighted mean of all other measured economies by \NpooledOne{} percent and the template by \Ntemplate{} percent (Table~\ref{tab:combos}). Choosing similar economies adds nothing. The country-specific component is idiosyncratic and is not shared within groups of similar economies. We cannot say what it reflects. Differences in product mix within industries, in the boundaries of firms and in long-lived supplier relations are natural candidates.

The premium is spread across the table. Cells off the diagonal carry \NoffDiagonalShare{} percent of it at every horizon. By buying industry, manufacturing accounts for \NmanufacturingShare{} percent of the one-year premium and transport and storage for \NtransportShare{} percent, and each of the eleven industry groups contributes less at ten years than at one (Table~\ref{tab:buyers}). Differences between economies account for \NvarCountryMeasured{} percent of the squared variation of kernels in this panel between 2000 and 2020.

\begin{figure}[!htbp]\centering
\includegraphics[width=\linewidth]{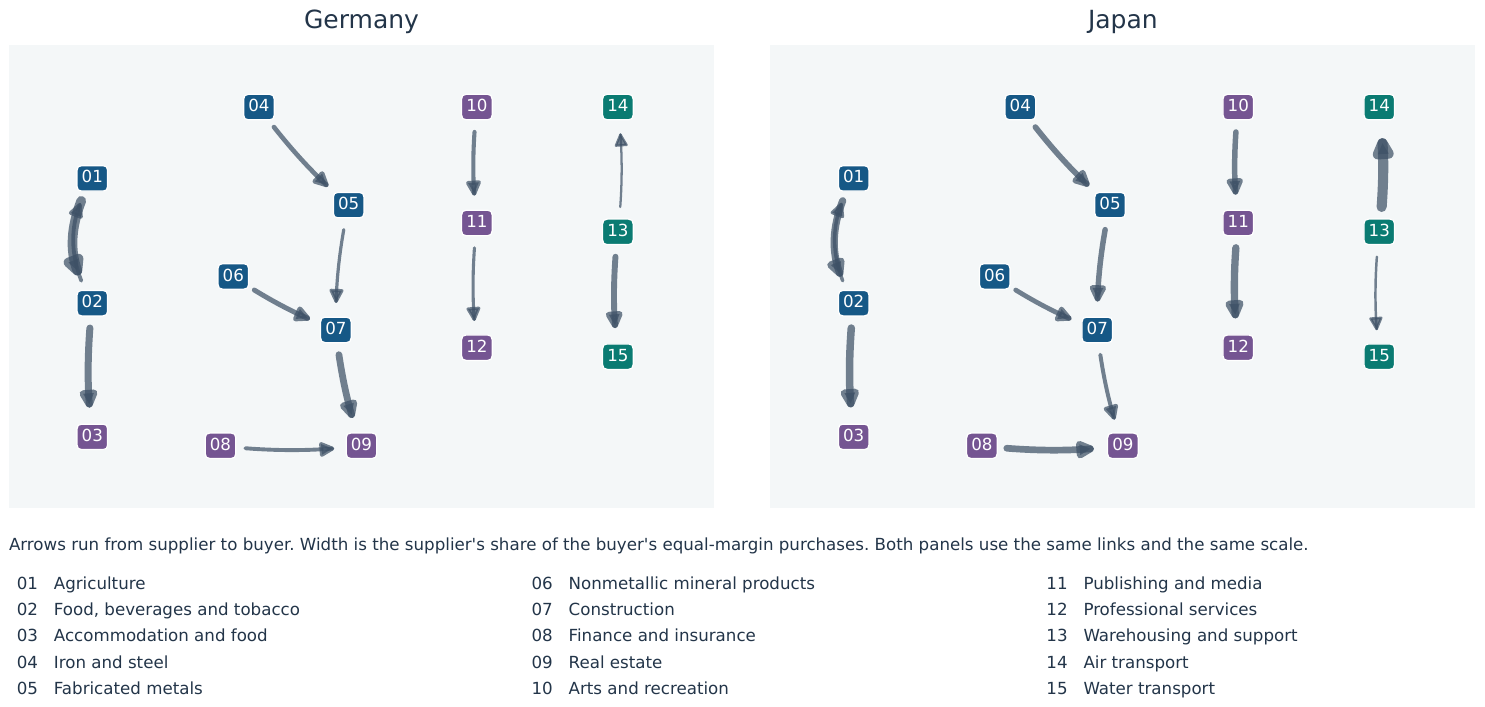}
\caption{Equal-margin networks of Germany and Japan. Nodes are industries in the 34-industry sample, and arrows run from the supplier to the buyer. The twelve arrows are the largest off-diagonal cells in the average of the two kernels over 2010 to 2019, years in which both economies have a listed national table. Both panels use those cells and the same width scale. Width is the supplier's share of the buyer's purchases in the equal-margin network. Own-industry purchases are omitted.}
\label{fig:network}
\end{figure}

Figure~\ref{fig:network} shows the links behind that variation for two economies in the panel. Germany and Japan have a listed national table in every year from 2010 to 2019. The twelve arrows are the largest cells in the average of their equal-margin networks over those years, and both panels draw the same cells. The food chain and the metals and construction chain are thick in both countries. Agriculture supplies \NnetDeAgFood{} percent of food's equal-margin purchases in Germany and \NnetJpAgFood{} percent in Japan, and nonmetallic mineral products supply \NnetDeMinerals{} percent of construction's equal-margin purchases in Germany and \NnetJpMinerals{} percent in Japan. The largest gap is in transport. Warehousing and transport support supplies \NnetDeAir{} percent of air transport's equal-margin purchases in Germany and \NnetJpAir{} percent in Japan, and \NnetDeWater{} percent of water transport's equal-margin purchases in Germany and \NnetJpWater{} percent in Japan. Construction supplies \NnetDeEstate{} percent of real estate's equal-margin purchases in Germany and \NnetJpEstate{} percent in Japan. A foreign history has to average over differences of this size.

Because peers contribute a stable common component, some weight on them could pay at long horizons. It does, modestly. Moving the latest table a quarter of the way toward the peer average raises the loss at horizons up to three years, makes no difference at five, and lowers the loss by \NshrinkGainEight{} points at eight years and by \NshrinkGainTen{} points at ten [\NshrinkGainTenLow, \NshrinkGainTenHigh] (Table~\ref{tab:combos}).

\subsection{Robustness}
The profile does not depend on details of the design (Table~\ref{tab:robust}). We drop the one panel economy whose kernel is template-like in some year, draw peers from all 80 economies, end the targets in 2019, and remove transport, utilities and public administration before rescaling. Across these variants the one-year premium lies between \NrobustLowOne{} and \NrobustHighOne{} points and the ten-year premium between \NrobustLowTen{} and \NrobustHighTen{}.

The equal weighting of industries is not responsible either. The equality in Proposition~\ref{prop:kernel}(b) uses uniform totals. For the recorded, uneven totals of the target year we fit each forecast kernel by RAS to those totals and compare the result with the recorded transaction shares, which weights cells by their value. The latest table then misallocates \NrasLatestOne{} percent of transaction value after one year, \NrasLatestFive{} percent after five and \NrasLatestTen{} percent after ten, against \NrasPeers{} percent for the peer kernel (Table~\ref{tab:ras}).

\subsection{Two databases}
\begin{figure}[!htbp]\centering
\includegraphics[width=\linewidth]{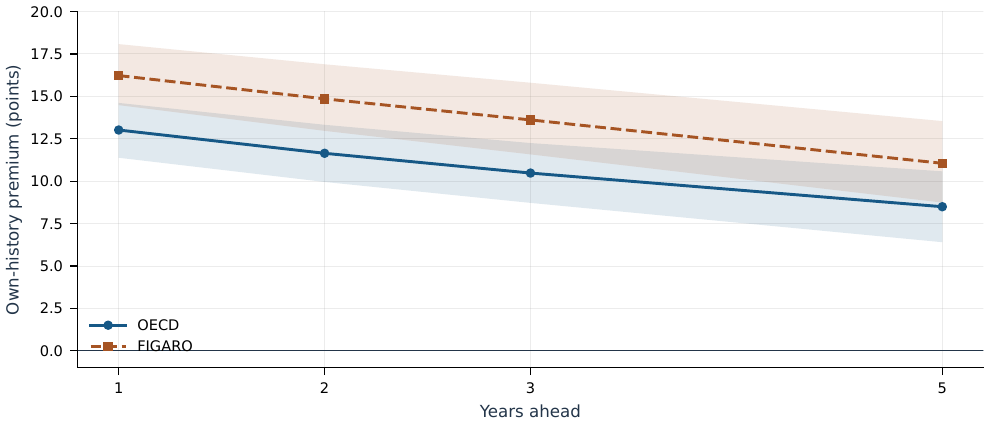}
\caption{Own-history premium in two databases for the \NnBoth{} economies with national tables in both from 2010 to 2020. Target years are 2015 to 2020, on 33 common industries. Each line trains and scores the forecasts in one database. Bands are 95 percent intervals from a bootstrap over economies.}
\label{fig:databases}
\end{figure}

The same experiment can be run in FIGARO for 2010 to 2022. Figure~\ref{fig:databases} shows the premium in each database for the \NnBoth{} economies with national tables in both, with targets from 2015 to 2020. One year ahead the latest table misplaces \NbothOECDLatestOne{} percent of the kernel in the OECD tables and \NbothFIGAROLatestOne{} percent in FIGARO. Five years ahead the figures are \NbothOECDLatestFive{} and \NbothFIGAROLatestFive{} percent. The premium declines from \NbothOECDGainOne{} to \NbothOECDGainFive{} points in the OECD tables and from \NbothFIGAROGainOne{} to \NbothFIGAROGainFive{} points in FIGARO. The FIGARO premium is about three points higher at each horizon because FIGARO kernels differ more across economies: the peer error is \NbothFIGAROPeersOne{} percent, against \NbothOECDPeersOne{} percent in the OECD tables.

The two databases are separate compilations of overlapping national sources. Their kernels for the same economy and year differ by \NcrossPanel{} percent on average, as much as a kernel differs from itself after about \NkernelAge{} years. Country-specific structure nevertheless carries across them. Own histories taken from the OECD tables forecast the FIGARO kernel of the same economy better than the OECD histories of its peers do, by \NcrossOFGainOne{} points one year ahead and in all \NcrossOFPositiveOne{} economies. In the other direction the premium is \NcrossFOGainOne{} points and positive in \NcrossFOPositiveOne{} economies (Table~\ref{tab:matched}).

\section{Country-Years Without a National Table}\label{sec:template}
\subsection{A common template in the OECD tables}
\begin{figure}[!htbp]\centering
\includegraphics[width=\linewidth]{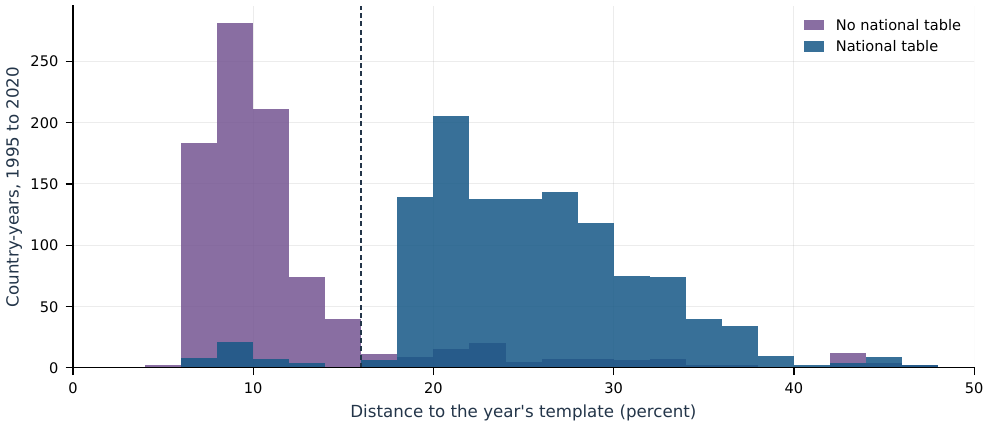}
\caption{Distance from each kernel to the template of its year, 1995 to 2020, by whether the OECD lists a national table. The dashed line marks 16 percent.}
\label{fig:template}
\end{figure}

We now turn to the country-years for which the OECD lists no national table. For each year we compute the average kernel of the economies without a national table and then update the set to the economies within 16 percent of that average until the set stops changing. We call the resulting average the template of that year and measure each economy's distance to it, leaving the economy itself out of the average. A country-year is template-like if its kernel lies within 16 percent of the template and specific otherwise. The threshold is the trough of the distribution in Figure~\ref{fig:template}. Appendix~\ref{app:template} reports thresholds from 12 to 20 percent and a version in which the set is not updated.

The distribution has two modes that line up with the documentation (Table~\ref{tab:template}, panel A). Between 1995 and 2020, country-years with a national table lie \NdistTable{} percent from the template on average, and \NshareLikeTable{} percent of them fall within the threshold. Country-years without a table lie \NdistNoTable{} percent away, and \NshareLikeNoTable{} percent fall within it. With economy and year effects, having a national table moves a kernel \NfeTable{} points away from the template [\NfeTableLow, \NfeTableHigh]. Two unmeasured economies in the same year are about \NtemplatePairNineteen{} percent apart (Table~\ref{tab:distance}), which is how far a measured economy is from its own kernel of four to five years earlier.

\begin{table}[!htbp]\centering\small
\caption{National tables and the template in the OECD tables}\label{tab:template}
\begin{tabular}{llrrrr}\toprule
\multicolumn{6}{l}{A. Distance to the template of the year}\\[2pt]
Years & Documentation & Country-years & Mean & Median & Template-like (\%) \\
\midrule
1995--2020 & National table & 1,180 & 25.4 & 25.0 & 3.6 \\
1995--2020 & No national table & 900 & 11.9 & 9.9 & 87.9 \\
2021 & National table & 37 & 18.5 & 19.8 & 37.8 \\
2021 & No national table & 43 & 12.2 & 10.8 & 86.0 \\
2022 & National table & 20 & 14.7 & 9.5 & 65.0 \\
2022 & No national table & 60 & 11.8 & 10.1 & 90.0 \\

\bottomrule\end{tabular}

\medskip
\begin{tabular}{lrrr}\toprule
\multicolumn{4}{l}{B. Change in the kernel from the previous year, 1996 to 2020}\\[2pt]
Documentation for the two years & Country-years & Mean & Median \\
\midrule
Table in both years & 992 & 7.0 & 4.5 \\
Table in neither year & 707 & 3.8 & 2.5 \\
Table arrives & 158 & 23.5 & 24.9 \\
Table lapses & 143 & 23.5 & 25.4 \\

\bottomrule\end{tabular}
\tabnote{80 economies and 34 industries. Distances and changes are total variation in percent. A country-year is template-like if its kernel lies within 16 percent of the template. Documentation is Table 3 of \citet{oecdnotes2025}.}
\end{table}

\begin{figure}[!htbp]\centering
\includegraphics[width=\linewidth]{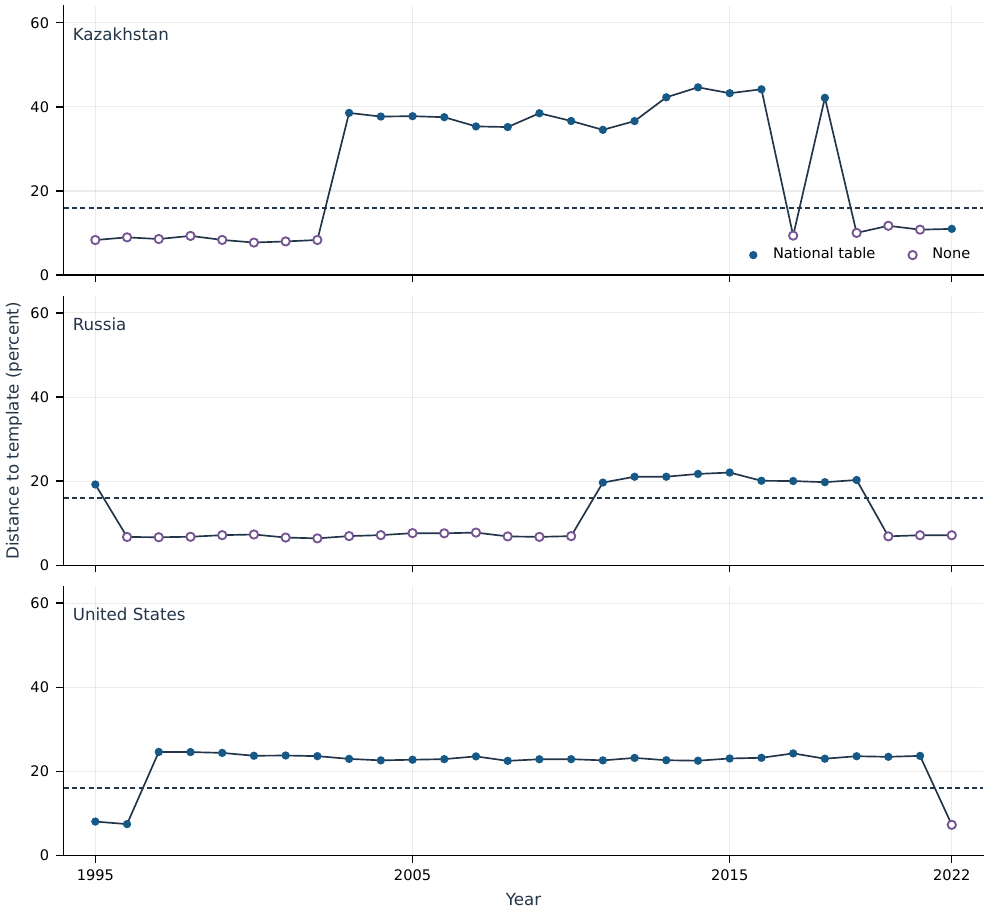}
\caption{Distance to the template for three economies. Filled points are years with a listed national table, and open points are years without one. The dashed line marks 16 percent.}
\label{fig:paths}
\end{figure}

Individual economies show the same thing over time (Figure~\ref{fig:paths}). Kazakhstan has a national table from 2003 to 2016 and again in 2018, and through 2021 its kernel leaves the template in exactly those years. Russia has national tables for 1995 and for 2011 to 2019 and is template-like in every other year. The United States is specific from 1997 to 2021 and template-like in 2022, the year of the construction cell in the introduction.

Panel B of Table~\ref{tab:template} measures what this does to apparent structural change. Between two years with national tables the kernel changes by \NchangeTableTable{} percent on average. Between two years without tables it changes by \NchangeNoneNone{} percent. When a table arrives or lapses it changes by \NchangeNoneTable{} percent, \NratioArrive{} times the movement of a normal year.

The pattern matches one of the two procedures in the OECD's methodology, the use of ``average production structures of other countries'' as initial values where sectoral detail is missing. A common initial structure fitted to each economy's own totals yields tables that look different in levels and in shares and yet have nearly the same kernel. That is what Proposition~\ref{prop:kernel}(a) implies and what we observe. We do not observe the compilation itself, and the correspondence with the documentation is not exact. About one in eight country-years without a table keeps a specific kernel. Australia before 2012 is an example: its off-benchmark years appear to be interpolated between national tables. In the other direction, \NshareLikeTable{} percent of the country-years listed with a table are template-like.

\begin{figure}[!htbp]\centering
\includegraphics[width=\linewidth]{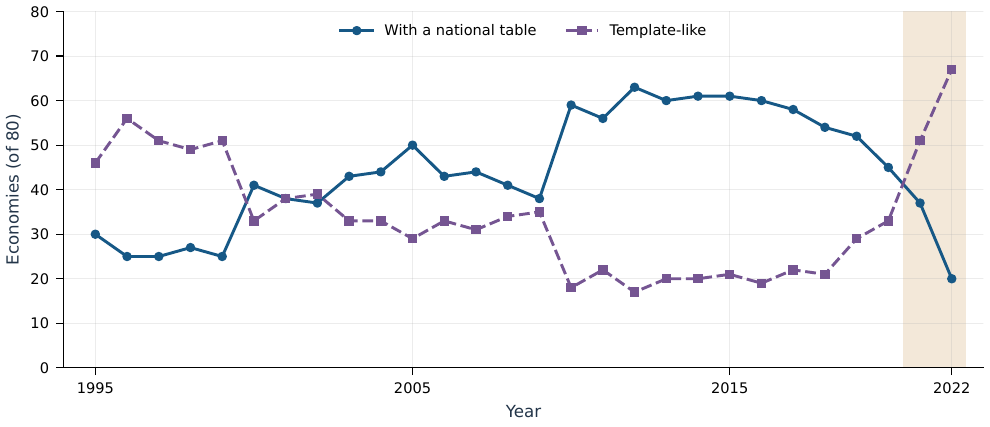}
\caption{Economies with a listed national table and economies within 16 percent of the template. Shading marks 2021 and 2022.}
\label{fig:counts}
\end{figure}

The last two years of the release deserve separate mention. Of the economies listed with a national table, \NshareLikeTableTwentyOne{} percent are template-like in 2021 and \NshareLikeTableTwentyTwo{} percent in 2022, against \NshareLikeTable{} percent before. By 2022, \NnLikeTwentyTwo{} of the 80 economies are template-like (Figure~\ref{fig:counts}). A listing records that a national source was available. It does not record which cells were taken from that source. A listed table is not an error-free observation of the production network, and a year without a listing need not have been filled by one common template.

\subsection{A different rule in FIGARO}
Eurostat's metadata describe a different rule for non-EU partners. Where no table is available, supply and use tables are estimated ``using SUTs from other years''. The kernels show what this means. Between two consecutive Eurostat estimates the kernel changes by \NfigaroChangeEstimated{} percent on average and by \NfigaroMedianEstimated{} percent at the median, against \NfigaroChangeDelivered{} and \NfigaroMedianDelivered{} percent between two tables delivered by national institutes (Table~\ref{tab:figaro}). The estimated tables keep the economy's own structure almost unchanged. FIGARO kernels lie within the threshold of the OECD template in \NfigaroNearTemplate{} percent of country-years.

\subsection{Which rule is closer to the recorded table}
The two compilers thus fill the same gap in opposite ways. One replaces an economy's structure with a cross-country average, and the other holds the economy's structure fixed. The experiment in Section~\ref{sec:persistence} compares both rules with the recorded kernel of economies that have a listed national table every year. That kernel is the outcome in the database, not an error-free observation of the production network. For these economies the template of the target year misses the recorded kernel by \Ntemplate{} percent (Table~\ref{tab:horizon}), about the same as the five nearest peers. The economy's own table misses by \NlatestOne{} percent when it is one year old, by \NlatestFive{} percent at five years and by \NlatestTen{} percent at ten. Even at ten years the own table is closer by \NtemplateMinusLatestTen{} points [\NtemplateMinusLatestTenLow, \NtemplateMinusLatestTenHigh], in \NtemplateMinusLatestTenPositive{} of \NnPanel{} economies. Figure~\ref{fig:long} shows the two curves meeting near fifteen years. The meeting point is approximate, and the template uses other economies' data from the target year. The ranking is the same after fitting to the recorded target-year totals: the template misallocates \NrasTemplate{} percent of transaction value, a ten-year-old own table \NrasLatestTen{} percent.

Measured economies might differ from the ones the template is applied to, so the arrival of tables provides a second check. When a national table follows a year without one, the kernel moves by \NchangeNoneTable{} percent on average. A normal year accounts for about \NchangeTableTable{} points of that, which puts the template error for the economies concerned at roughly \NtemplateErrorBound{} points or more.

Each rule has a cost. Holding structure fixed understates change by construction. Substituting a template overstates similarity across economies and creates a break whenever a table arrives or lapses. For the kernel, in this sample, carrying a structure forward has the smaller error on the short side of that approximate crossing.

\section{What the Template Does to Measured Change}\label{sec:change}
\subsection{Convergence after 2019}
\begin{figure}[!htbp]\centering
\includegraphics[width=\linewidth]{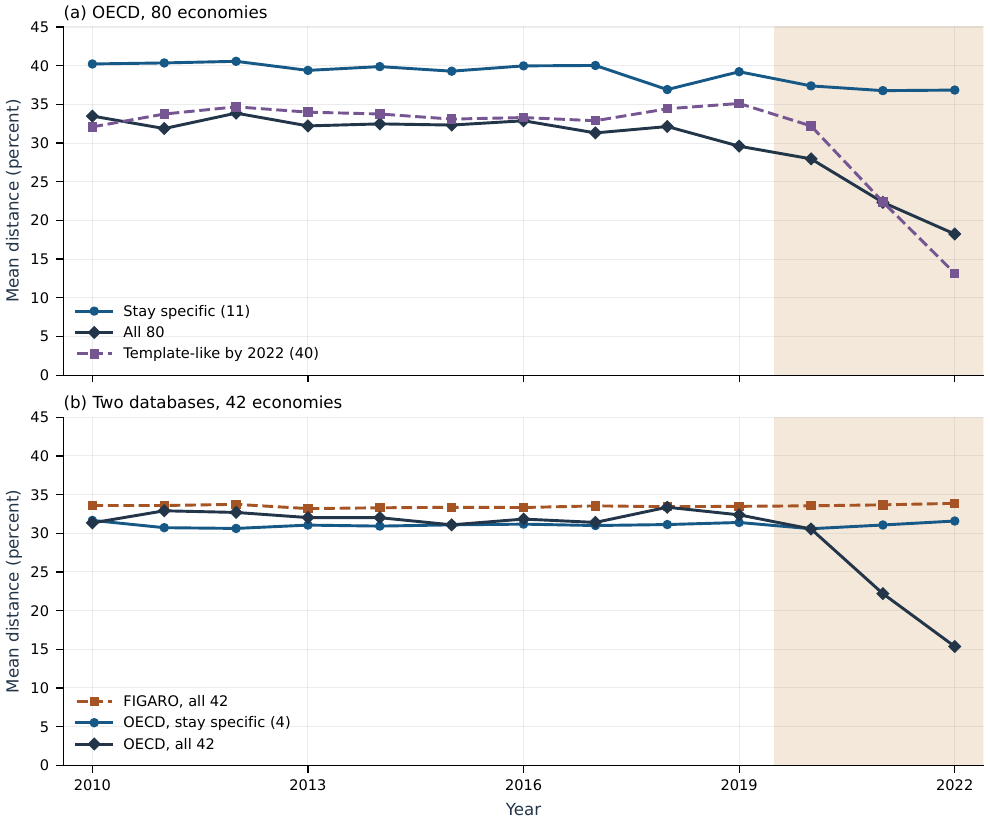}
\caption{Mean distance between economies, 2010 to 2022. Panel (a) uses the OECD tables for 80 economies and separates the economies whose kernels are specific in both 2019 and 2022 from those that are specific in 2019 and template-like in 2022. Panel (b) uses the 42 economies and 33 industries available in both databases. Shading marks 2020 to 2022.}
\label{fig:convergence}
\end{figure}

Read at face value, the OECD tables say that national production structures converged sharply after 2019. Table~\ref{tab:distance} and Figure~\ref{fig:convergence} show where the convergence comes from. Average distance between the kernels of all 80 economies falls from \NdistAllNineteen{} percent in 2019 to \NdistAllTwentyTwo{} percent in 2022. Among the \NnSwitch{} economies that are specific in 2019 and template-like in 2022, it falls from \NdistSwitchNineteen{} to \NdistSwitchTwentyTwo{} percent. Among the \NnStay{} economies that are specific in both years it moves from \NdistStayNineteen{} to \NdistStayTwentyTwo{} percent, about where it stood in 2018 (\NdistStayEighteen{} percent).

\begin{table}[!htbp]\centering\small
\caption{Mean distance between the kernels of different economies}\label{tab:distance}
\begin{tabular}{lrrrrr}\toprule
Kernels compared & Economies & Pairs & 2019 & 2022 & Change \\
\midrule
OECD, all 80 economies & 80 & 3,160 & 29.6 & 18.2 & \ensuremath{-}11.4 \\
\quad specific in 2019 and 2022 & 11 & 55 & 39.2 & 36.8 & \ensuremath{-}2.4 \\
\quad specific in 2019, template-like in 2022 & 40 & 780 & 35.1 & 13.2 & \ensuremath{-}22.0 \\
\quad template-like in 2019 and 2022 & 27 & 351 & 13.9 & 14.9 & 1.1 \\
\addlinespace
OECD, 42 matched economies & 42 & 861 & 32.4 & 15.4 & \ensuremath{-}17.0 \\
\quad specific in 2019 and 2022 & 4 & 6 & 31.4 & 31.6 & 0.2 \\
FIGARO, 42 matched economies & 42 & 861 & 33.5 & 33.9 & 0.4 \\
\quad the same four economies & 4 & 6 & 29.8 & 31.1 & 1.2 \\

\bottomrule\end{tabular}
\tabnote{Total variation distance in percent, averaged over all pairs of distinct economies in the group. The upper block uses 80 economies and 34 industries, the lower block 42 economies and 33 industries available in both databases. Specific and template-like refer to the OECD kernel of the economy in the stated year.}
\end{table}

The matched sample gives the same answer from the other database. Among the 42 economies the OECD distance falls from \NmOECDAllNineteen{} to \NmOECDAllTwentyTwo{} percent, while the FIGARO distance is \NmFIGAROAllNineteen{} percent in 2019 and \NmFIGAROAllTwentyTwo{} percent in 2022. Four of the 42 economies have a specific OECD kernel in both 2019 and 2022: Czechia, Japan, the Netherlands and the United Kingdom. Their mutual distance is \NmOECDStayNineteen{} percent in 2019 and \NmOECDStayTwentyTwo{} percent in 2022 in the OECD tables.

The fall is visible in weaker form before any rescaling. Among the 42 matched economies the OECD distance falls from \NoldTransNineteen{} to \NoldTransTwentyTwo{} percent in raw transaction shares and from \NoldBuyerNineteen{} to \NoldBuyerTwentyTwo{} percent in buyer shares, while the FIGARO figures rise slightly. Industry sizes and input intensities are country specific in every year, including the unmeasured ones, and they dilute the common kernel in these statistics.

\subsection{Pooled forecast evaluations}
\begin{figure}[!htbp]\centering
\includegraphics[width=\linewidth]{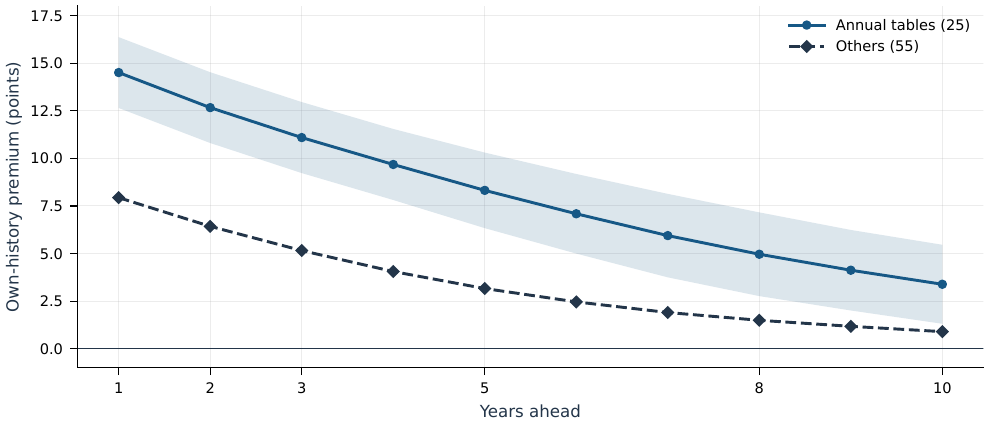}
\caption{Own-history premium by horizon. The solid line is the \NnPanel{} economies with a national table every year, with its two-way clustered interval. The dashed line is the other \NnRest{} economies, with peers drawn from all 80.}
\label{fig:premium}
\end{figure}

A forecast evaluation that pools measured and unmeasured country-years gives a misleading account of persistence. For the \NnRest{} economies outside the measured panel the premium is \NgainOtherOne{} points at one year and \NgainOtherTen{} points at ten (Figure~\ref{fig:premium}).

 These economies move in and out of the template, which their own history cannot predict, and their nearest peers are often other template-like economies with nearly identical kernels. Pooling all 80 economies and extending the targets to 2022, as one would without the documentation, gives \NoldGainOne{} points at one year and \NoldGainTen{} points at ten, with the own history ahead in \NoldPositiveTen{} of 80 economies at ten years.

\begin{figure}[!htbp]\centering
\includegraphics[width=\linewidth]{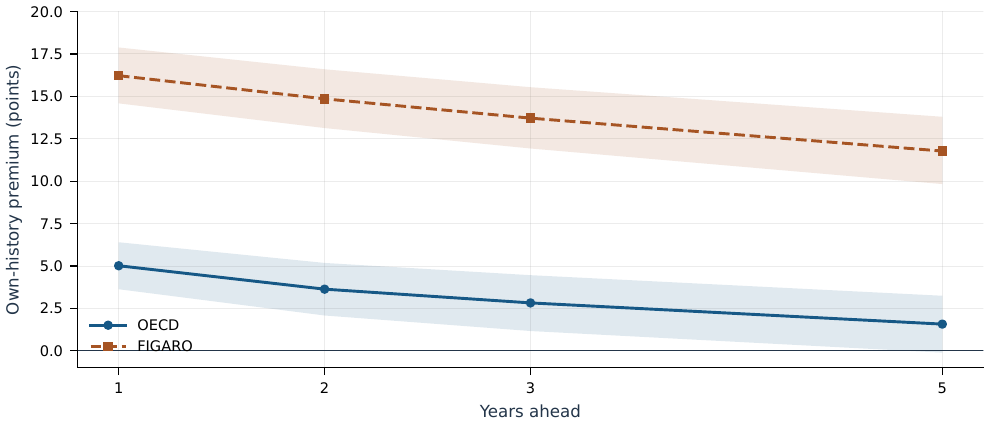}
\caption{Own-history premium in two databases for all 42 matched economies, with target years 2019 to 2022. Each line trains and scores the forecasts in one database on 33 common industries. Bands are 95 percent intervals from a bootstrap over economies.}
\label{fig:databases-late}
\end{figure}

The comparison of databases is distorted in the same way. With all 42 matched economies and targets from 2019 to 2022, the one-year premium is \NallOECDGainOne{} points in the OECD tables and \NallFIGAROGainOne{} points in FIGARO (Figure~\ref{fig:databases-late}), a gap that would suggest the two compilers disagree about persistence itself.

 On the economies measured in both, the corresponding figures are \NbothOECDGainOne{} and \NbothFIGAROGainOne{} points.

\subsection{Multipliers and supplier shares}
What does the template do to the statistics that users compute? We consider two, both on the matched sample. The domestic output multiplier of an industry is the column sum of $(I-A)^{-1}$, where $A_{ij}=Z_{ij}/x_j$ and $x_j$ is the buyer's gross output. It measures the output generated within the retained domestic system by one unit of final demand. A buyer's supplier shares are its purchases from each other domestic industry, divided by its purchases from all other domestic industries.

\begin{figure}[!htbp]\centering
\includegraphics[width=\linewidth]{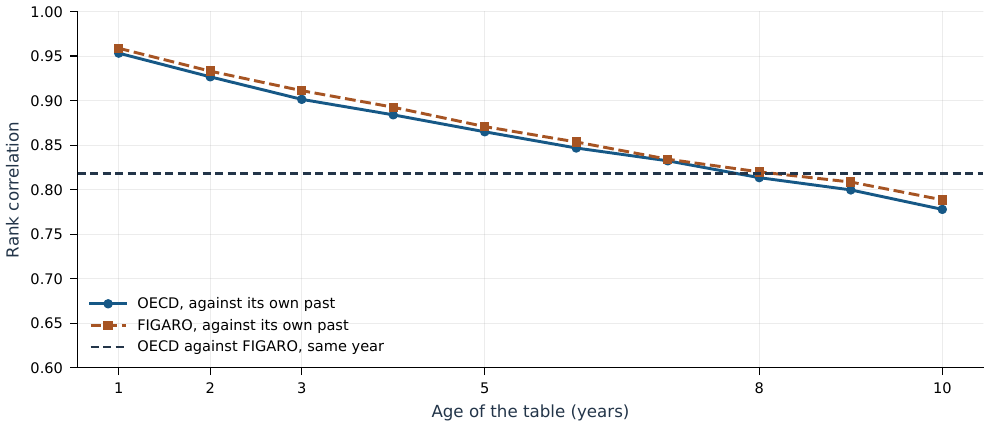}
\caption{Rank correlation between a database's multipliers and its own multipliers from earlier years, for the \NnBoth{} economies measured in both databases from 2010 to 2020. The dashed line is the correlation between the two databases in the same year.}
\label{fig:mult-age}
\end{figure}

Agreement between the databases provides a yardstick. For the \NnBoth{} economies measured in both, the rank correlation between OECD and FIGARO multipliers within an economy averages \NrhoCrossPanel{} between 2010 and 2020. The correlation between a table's multipliers and those of the same database eight years earlier is \NrhoLagEight{} (Figure~\ref{fig:mult-age}). Choosing between the two databases changes sector rankings about as much as using a table of that age.

\begin{figure}[!htbp]\centering
\includegraphics[width=\linewidth]{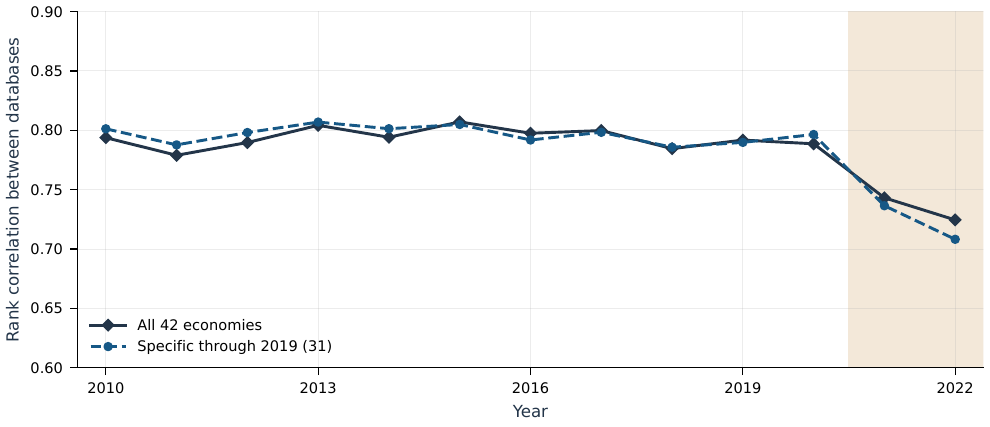}
\caption{Mean within-economy rank correlation between OECD and FIGARO domestic output multipliers. The dashed series is the \NnCleanMult{} economies whose OECD kernels are specific in every year from 2010 to 2019. Shading marks 2021 and 2022.}
\label{fig:mult-year}
\end{figure}

After 2020 agreement falls. Across all 42 economies the mean rank correlation lies between \NrhoEarlyMin{} and \NrhoEarlyMax{} in every year from 2010 to 2020, and it is \NrhoTwentyOne{} in 2021 and \NrhoTwentyTwo{} in 2022 (Figure~\ref{fig:mult-year}).

 Mean multipliers remain almost identical, at \NmultOECD{} and \NmultFIGARO{} in 2022.

Table~\ref{tab:events} relates the fall to the template. We compare the change in each statistic in the year an economy's OECD table becomes template-like with the change in the same year for economies whose tables stay specific, pooling 2020, 2021 and 2022. The distance between the two databases' kernels rises by \NeventKernelSwitch{} points for the switchers and not at all for the others. The distance between their supplier shares rises by \NeventSupplierSwitch{} points against \NeventSupplierControl{}, a difference of \NeventSupplierDiff{} points [\NeventSupplierLow, \NeventSupplierHigh]. The rank correlation of multipliers changes by \NeventRhoSwitch{} for the switchers and by \NeventRhoControl{} for the others, a difference of \NeventRhoDiff{} [\NeventRhoLow, \NeventRhoHigh]. The gap in domestic input intensity, the column sums of $A$, does not respond: the difference is \NeventIntensityDiff{} [\NeventIntensityLow, \NeventIntensityHigh].

\begin{table}[!htbp]\centering\small
\caption{Agreement between databases in the year a table becomes template-like}\label{tab:events}
\setlength{\tabcolsep}{5pt}
\begin{tabular}{lrrrl}\toprule
 & \multicolumn{2}{c}{Change from the previous year} & & \\
\cmidrule(lr){2-3}
Statistic compared across databases & Switches & Stays specific & Difference & 95\% interval \\
\midrule
Kernel distance between databases & 9.0 & 0.0 & 9.0 & [5.4, 12.2] \\
Supplier shares, distance between databases & 7.2 & 0.1 & 7.1 & [3.7, 10.1] \\
Multiplier ranking, rank correlation & \ensuremath{-}0.061 & \ensuremath{-}0.006 & \ensuremath{-}0.055 & [\ensuremath{-}0.094, \ensuremath{-}0.013] \\
Multiplier, mean absolute gap & 0.013 & 0.007 & 0.007 & [\ensuremath{-}0.004, 0.014] \\
Domestic input intensity, mean absolute gap & 0.006 & 0.003 & 0.003 & [\ensuremath{-}0.004, 0.008] \\

\bottomrule\end{tabular}
\tabnote{42 matched economies and 33 industries. A switch is a year in 2020 to 2022 in which the OECD kernel of an economy is template-like after being specific the year before (\NnSwitchEvents{} switches). The comparison group is the economies whose OECD kernels are specific in both years (\NnControlObs{} observations). Each statistic measures the difference between the OECD and FIGARO tables of the same economy and year. Kernel and supplier-share distances are total variation in percent. Intervals resample economies.}
\end{table}

The template thus changes who supplies whom and leaves how much each industry buys to the national accounts totals. Multiplier levels depend mostly on the second and barely move. Rankings and supplier-specific exposures depend on the first.

\section{Conclusion}\label{sec:conclusion}
National input-output tables contain a large country-specific component that lasts. In economies with annual national tables, last year's table describes this year's interindustry pattern with about a quarter of the error of the tables of the most similar economies. In this sample the error of an own table and the distance to that year's template meet near fifteen years. The crossing uses few origins beyond ten years and a template built from other economies in the target year, so it describes this period. It is not a shelf life, and it is not a rule for real-time updating. Two separately compiled databases show similar profiles for the economies with listed tables in both. Many of the apparent exceptions come from country-years with no listed national table. In the OECD tables most of those kernels lie near a common template, so that economies seem to converge when listed coverage thins and to change abruptly when a national table arrives.

The most recent years of a release are the ones most in demand, for instance to study how production adjusted after 2020, and they are the years with the fewest listed national tables. For users of multi-country tables the implication is to separate country-years by the compilers' availability list before interpreting changes in structure. Distance from the template, which is initialized from that list, flags further country-years whose kernels look like the common pattern, including many in the two most recent years. A listing that a national table was available does not mean every cell was taken from it, and a year without a listing was not necessarily filled by one cross-country template. Analyses of supplier-specific exposure, of structural similarity between economies and of changes in sector rankings are the ones at risk. Aggregate multipliers and statistics driven by industry size and input intensity are much less affected. Where the comparison can be scored, carrying forward an economy's own structure has a smaller kernel error than a cross-country average over the ten years the forecasts cover, and a flag for each estimated country-year would make the distinction checkable. For a researcher describing a year with no listed table, the most recent national table rescaled to current totals has the smaller error inside that ten-year window. We do not read the fifteen-year crossing as a prescription.

Several limits apply. The kernel describes domestic sourcing and leaves imported inputs aside. The 34 industries cover about three quarters of domestic intermediate transactions and exclude mining, refining, machinery and motor vehicles. The measured panel is mostly European and high-income. National tables between benchmark years contain estimation of their own. We use the recorded table as the outcome, not as an error-free observation of technology. All tables come from a single release of each database, so the experiment describes how well earlier reference years predict later ones in that release and says nothing about what a forecaster would have had in real time. Other releases and other multi-regional databases can be checked in the same way. Finally, our reading of how unlisted years are filled rests on published documentation and on the pattern of the kernels. We have not seen the compilers' working files.

\section*{Data and code availability}
The replication files contain the code that produces every number, table and figure in the paper, the extraction of the domestic blocks from the published OECD and Eurostat tables, and the two documentation tables used to classify country-years.

\appendix
\setcounter{table}{0}
\setcounter{figure}{0}
\setcounter{equation}{0}
\renewcommand{\thetable}{A\arabic{table}}
\renewcommand{\thefigure}{A\arabic{figure}}
\renewcommand{\theequation}{A\arabic{equation}}

\section{Data}\label{app:data}
\paragraph{Industries.} The OECD tables have 50 industries. A kernel exists only if every industry has positive domestic sales and purchases, and comparisons across economies and years require a common set. The 34 industries that meet the condition in every economy and year from 1995 to 2004 also meet it in every later year. The 16 excluded industries are forestry, fishing, the five mining industries, refining, rubber and plastics, electronics, electrical equipment, machinery, motor vehicles, the two other transport equipment industries, and households as employers. Between 1995 and 2004 motor vehicles fail the condition in ten country-years and machinery in one. Table~\ref{tab:coverage} reports how much of each economy's domestic intermediate transactions the retained industries cover. Across country-years the fifth and ninety-fifth percentiles of the share with both industries retained are 56.9 and 92.5 percent.

\begin{table}[!htbp]\centering\small
\caption{Share of domestic intermediate transactions covered}\label{tab:coverage}
\begin{tabular}{lrrrrr}\toprule
Industries & Buyer retained & Supplier retained & Both retained & One retained & Neither \\
\midrule
OECD core 34 & 84.3 & 83.8 & 75.8 & 16.5 & 7.7 \\
OECD reduced 26 & 70.7 & 69.2 & 56.7 & 26.4 & 16.9 \\
OECD matched 33 & 86.4 & 87.5 & 79.4 & 15.0 & 5.5 \\
FIGARO matched 33 & 87.3 & 88.1 & 80.4 & 14.6 & 5.0 \\

\bottomrule\end{tabular}
\tabnote{Percent of nominal domestic intermediate transactions in the native tables, averaged over country-years. The native tables have 50 industries in the OECD source and 64 in FIGARO. The 26-industry row removes transport, utilities and public administration from the 34.}
\end{table}

\paragraph{Matched sample.} FIGARO has 64 industries. We aggregate transactions and gross output in both sources to 33 groups that can be formed from whole industries in each, for the 42 economies present in both, from 2010 to 2022. Aggregation is applied to transactions before any ratio is taken. The OECD tables are in US dollars and FIGARO in euros. The unit does not affect kernels, shares or multipliers.

\paragraph{Documentation.} Table 3 of \citet{oecdnotes2025} codes each economy and year as having no national table, supply and use tables only, input-output tables only, or both. We treat the last three as a national table. Table 1 in section 18.3 of \citet{eurostatmeta} gives the origin of the supply and use tables of non-EU partners by year. Both tables were retrieved on 18 November 2026 and are included with the replication files.

\paragraph{Panels.} The \NnPanel{} economies with an OECD-listed national table in every year from 2000 to 2020 are \NpanelList. One of them (\NpanelDropped) has a template-like kernel in one year. The \NnBoth{} economies with national tables in both databases from 2010 to 2020 are \NbothList. Three of them (\NbothDropped) have a template-like OECD kernel in some year. The \NnStay{} economies with specific OECD kernels in both 2019 and 2022 are \NstayList.

\section{Proof of Proposition~\ref{prop:kernel}}\label{app:proof}
(a) Let $Q(Z)=D^rZD^c$. Then $(D^rR^{-1})(RZS)(S^{-1}D^c)=Q(Z)$ is a diagonal scaling of $RZS$ with uniform row and column totals, and uniqueness of the kernel gives $Q(RZS)=Q(Z)$.

(b) The biproportional adjustment of $Z$ to given totals is the unique matrix of the form $RZS$ with those totals \citep{bacharach1970}. If $Q(Z)=Q(\tilde Z)$, then $D^rZD^c=\tilde D^r\tilde Z\tilde D^c$, so $\tilde Z=(\tilde D^r)^{-1}D^r\,Z\,D^c(\tilde D^c)^{-1}$ has the form $RZS$ and the totals of $\tilde Z$. It is therefore the adjustment. Conversely, if the adjustment equals $\tilde Z$ then $\tilde Z=RZS$ and (a) gives $Q(\tilde Z)=Q(Z)$. When the totals of $\tilde Z$ are uniform and its mass is one, $\tilde Z=Q(\tilde Z)$ and the adjustment of $Z$ is $Q(Z)$, so the error is $d(Q(Z),Q(\tilde Z))$.

(c) Buyer $j$ chooses quantities $m_{ij}$ to minimize $\sum_ip_im_{ij}$ subject to $\bigl(\sum_ia_{ij}^{1/\sigma}m_{ij}^{(\sigma-1)/\sigma}\bigr)^{\sigma/(\sigma-1)}=M_j$. The solution is $m_{ij}=a_{ij}(p_i/P_j)^{-\sigma}M_j$ with $P_j=\bigl(\sum_ia_{ij}p_i^{1-\sigma}\bigr)^{1/(1-\sigma)}$, so expenditure is $Z_{ij}=p_im_{ij}=p_i^{1-\sigma}\,a_{ij}\,P_j^{\sigma}M_j$. This is $RAS$ with $R=\operatorname{diag}(p_i^{1-\sigma})$ and $S=\operatorname{diag}(P_j^{\sigma}M_j)$, and (a) applies. The Cobb--Douglas case is the limit $\sigma\to1$, in which $R$ is the identity. With buyer-specific elasticities, $\log Z_{ij}=\log a_{ij}+(1-\bar\sigma)\log p_i-(\sigma_j-\bar\sigma)\log p_i+\log(P_j^{\sigma_j}M_j)$. The second and fourth terms are row and column factors, and the third is the interaction mentioned in the text.

\section{The Template}\label{app:template}
For year $t$ let $S_0$ be the set of economies without an OECD-listed national table. Given a set $S_k$, the distance of economy $c$ is $d\bigl(Q_{ct},\bar Q_{t}^{(-c)}\bigr)$, where $\bar Q_t^{(-c)}$ is the mean kernel of the economies in $S_k$ other than $c$. The next set $S_{k+1}$ contains the economies with a distance below 0.16. The iteration stops when the set no longer changes, which it does in every year, and the template is the mean kernel of the final set. The initial set is taken from the OECD availability list. Later rounds can add or drop economies, but they do not make the template independent of that list. From one year to the next the template itself moves by \NtemplateDrift{} percent on average.

Table~\ref{tab:thresholds} varies the threshold. Between 0.14 and 0.18 the share of table-backed country-years classified as template-like stays between 3.5 and 4.1 percent, and at 0.20 the threshold begins to cut into the second mode. The last row uses the distance to the mean kernel of $S_0$ without any update. That row, like the iterated definition, starts from the availability list. Table~\ref{tab:years} gives the counts by year.

\begin{table}[!htbp]\centering\small
\caption{Share of country-years classified as template-like under other thresholds}\label{tab:thresholds}
\begin{tabular}{lrrrr}\toprule
 & \multicolumn{2}{c}{1995 to 2020} & \multicolumn{2}{c}{2021 and 2022}\\
\cmidrule(lr){2-3}\cmidrule(lr){4-5}
Threshold & No national table & National table & No national table & National table \\
\midrule
0.12 & 75.2 & 3.1 & 64.1 & 43.9 \\
0.14 & 83.4 & 3.5 & 77.7 & 47.4 \\
0.16 (baseline) & 87.9 & 3.6 & 88.3 & 47.4 \\
0.18 & 89.1 & 4.1 & 91.3 & 50.9 \\
0.20 & 90.1 & 15.8 & 92.2 & 59.6 \\
0.16, documentary template & 87.3 & 3.6 & 88.3 & 47.4 \\

\bottomrule\end{tabular}
\tabnote{Percent of country-years whose kernel lies within the threshold of the template, by OECD-listed availability of a national table.}
\end{table}

\begin{table}[!htbp]\centering\small
\caption{National tables and template-like kernels by year}\label{tab:years}
\setlength{\tabcolsep}{4.5pt}
\begin{tabular}{rrrrr@{\hskip 18pt}rrrrr}\toprule
 & National & Template- & \multicolumn{2}{c}{of which} & & National & Template- & \multicolumn{2}{c}{of which}\\
Year & tables & like & no table & table & Year & tables & like & no table & table \\
\midrule
1995 & 30 & 46 & 43 & 3 & 2009 & 38 & 35 & 35 & 0 \\
1996 & 25 & 56 & 51 & 5 & 2010 & 59 & 18 & 17 & 1 \\
1997 & 25 & 51 & 51 & 0 & 2011 & 56 & 22 & 20 & 2 \\
1998 & 27 & 49 & 49 & 0 & 2012 & 63 & 17 & 15 & 2 \\
1999 & 25 & 51 & 50 & 1 & 2013 & 60 & 20 & 18 & 2 \\
2000 & 41 & 33 & 32 & 1 & 2014 & 61 & 20 & 17 & 3 \\
2001 & 38 & 38 & 38 & 0 & 2015 & 61 & 21 & 18 & 3 \\
2002 & 37 & 39 & 39 & 0 & 2016 & 60 & 19 & 16 & 3 \\
2003 & 43 & 33 & 33 & 0 & 2017 & 58 & 22 & 17 & 5 \\
2004 & 44 & 33 & 33 & 0 & 2018 & 54 & 21 & 20 & 1 \\
2005 & 50 & 29 & 27 & 2 & 2019 & 52 & 29 & 25 & 4 \\
2006 & 43 & 33 & 32 & 1 & 2020 & 45 & 33 & 30 & 3 \\
2007 & 44 & 31 & 31 & 0 & 2021 & 37 & 51 & 37 & 14 \\
2008 & 41 & 34 & 34 & 0 & 2022 & 20 & 67 & 54 & 13 \\

\bottomrule\end{tabular}
\tabnote{Number of economies out of 80. National tables are those listed in \citet{oecdnotes2025}.}
\end{table}

\section{Forecasts}\label{app:forecast}
\paragraph{Inference.} For a given horizon the premium is an array indexed by target year and economy. The economy bootstrap resamples whole columns 10,000 times. The two-way variance adds the cluster sums by economy and by target year, subtracts the sum over single observations, applies finite-cluster corrections and uses the smaller number of clusters minus one as degrees of freedom. The joint bootstrap resamples economies and circular blocks of three adjacent target years 4,000 times. Table~\ref{tab:intervals} reports all three for the measured panel, together with the comparison between the template and the latest own table.

\begin{table}[!htbp]\centering\small
\caption{Intervals for the measured panel}\label{tab:intervals}
\setlength{\tabcolsep}{4.5pt}
\begin{tabular}{rlllrlr}\toprule
 & \multicolumn{3}{c}{Own-history premium, 95\% interval} & \multicolumn{3}{c}{Template loss minus latest-table loss}\\
\cmidrule(lr){2-4}\cmidrule(lr){5-7}
Years ahead & Economies & Two-way & Economies and blocks & Mean & 95\% interval & Positive \\
\midrule
1 & [13.2, 15.9] & [12.6, 16.4] & [12.8, 16.4] & 17.8 & [15.5, 20.0] & 25/25 \\
2 & [11.2, 14.1] & [10.8, 14.5] & [10.9, 14.5] & 15.1 & [13.1, 17.2] & 25/25 \\
3 & [9.6, 12.5] & [9.2, 13.0] & [9.3, 12.9] & 13.2 & [11.0, 15.4] & 25/25 \\
5 & [6.8, 9.8] & [6.3, 10.3] & [6.6, 10.2] & 9.6 & [7.4, 11.8] & 24/25 \\
8 & [3.3, 6.6] & [2.8, 7.2] & [3.0, 6.9] & 5.2 & [2.6, 7.9] & 22/25 \\
10 & [1.7, 5.1] & [1.3, 5.5] & [1.6, 5.2] & 3.1 & [0.7, 5.5] & 19/25 \\

\bottomrule\end{tabular}
\tabnote{\NnPanel{} economies, target years 2014 to 2020, in points of total variation. The interval in the right block is two-way clustered.}
\end{table}

\begin{table}[!htbp]\centering\small
\caption{Own five-year mean and peers, by group of economies}\label{tab:groups}
\setlength{\tabcolsep}{3.6pt}
\begin{tabular}{rrrrrrrrrrrrr}\toprule
 & \multicolumn{4}{c}{National table every year} & \multicolumn{4}{c}{Other economies} & \multicolumn{4}{c}{All 80}\\
\cmidrule(lr){2-5}\cmidrule(lr){6-9}\cmidrule(lr){10-13}
Years & Own & Peers & Prem. & Pos. & Own & Peers & Prem. & Pos. & Own & Peers & Prem. & Pos. \\
\midrule
1 & 9.6 & 24.1 & 14.5 & 25/25 & 11.7 & 19.6 & 7.9 & 49/55 & 11.0 & 20.9 & 9.8 & 74/80 \\
2 & 11.5 & 24.1 & 12.7 & 25/25 & 13.3 & 19.7 & 6.4 & 48/55 & 12.7 & 21.0 & 8.2 & 73/80 \\
3 & 13.1 & 24.2 & 11.1 & 25/25 & 14.7 & 19.8 & 5.2 & 47/55 & 14.2 & 21.0 & 6.8 & 72/80 \\
5 & 16.1 & 24.4 & 8.3 & 24/25 & 16.9 & 20.0 & 3.2 & 42/55 & 16.6 & 21.2 & 4.6 & 66/80 \\
8 & 19.8 & 24.7 & 5.0 & 23/25 & 18.8 & 20.2 & 1.5 & 37/55 & 19.1 & 21.4 & 2.4 & 57/80 \\
10 & 21.6 & 25.0 & 3.4 & 20/25 & 19.5 & 20.4 & 0.9 & 37/55 & 20.1 & 21.6 & 1.4 & 56/80 \\

\bottomrule\end{tabular}
\tabnote{Target years 2014 to 2020. Peers are drawn from the measured panel in the first block and from all 80 economies in the other two. Losses in percent, premium in points, and the number of economies with a positive mean premium.}
\end{table}

\begin{table}[!htbp]\centering\small
\caption{Combinations of own and peer forecasts}\label{tab:combos}
\setlength{\tabcolsep}{4.5pt}
\begin{tabular}{rrrrrrrrrl}\toprule
 & \multicolumn{3}{c}{Latest own table} & \multicolumn{3}{c}{Own five-year mean} & & \multicolumn{2}{c}{Gain from 25\% on latest}\\
\cmidrule(lr){2-4}\cmidrule(lr){5-7}\cmidrule(lr){9-10}
Years & Alone & 25\% peers & 50\% peers & Alone & 25\% peers & 50\% peers & All others & Mean & 95\% interval \\
\midrule
1 & 6.6 & 9.1 & 13.6 & 9.6 & 11.4 & 15.0 & 24.7 & \ensuremath{-}2.5 & [\ensuremath{-}3.2, \ensuremath{-}1.9] \\
2 & 9.2 & 10.8 & 14.6 & 11.5 & 12.7 & 15.8 & 24.8 & \ensuremath{-}1.6 & [\ensuremath{-}2.3, \ensuremath{-}1.0] \\
3 & 11.1 & 12.2 & 15.4 & 13.1 & 13.9 & 16.5 & 24.8 & \ensuremath{-}1.1 & [\ensuremath{-}1.7, \ensuremath{-}0.4] \\
5 & 14.7 & 14.9 & 17.0 & 16.1 & 16.2 & 18.0 & 24.9 & \ensuremath{-}0.2 & [\ensuremath{-}0.8, 0.5] \\
8 & 19.1 & 18.3 & 19.3 & 19.8 & 19.1 & 19.9 & 25.1 & 0.8 & [0.2, 1.5] \\
10 & 21.2 & 20.0 & 20.5 & 21.6 & 20.5 & 20.9 & 25.4 & 1.2 & [0.6, 1.8] \\

\bottomrule\end{tabular}
\tabnote{\NnPanel{} economies, target years 2014 to 2020, losses in percent. Peers are the five nearest economies. All others is the unweighted mean of the five-year means of every other panel economy. The last two columns give the loss of the latest table minus the loss of the latest table moved a quarter of the way toward the peers, with a two-way clustered interval.}
\end{table}

\begin{table}[!htbp]\centering\small
\caption{Own-history premium under other designs}\label{tab:robust}
\begin{tabular}{lrrrrrrr}\toprule
 & & \multicolumn{6}{c}{Years ahead}\\
\cmidrule(lr){3-8}
Design & Economies & 1 & 2 & 3 & 5 & 8 & 10 \\
\midrule
Baseline & 25 & 14.5 & 12.7 & 11.1 & 8.3 & 5.0 & 3.4 \\
Never template-like, 2000--2020 & 24 & 14.5 & 12.6 & 11.0 & 8.1 & 4.7 & 3.2 \\
Peers drawn from all 80 economies & 25 & 14.0 & 12.1 & 10.5 & 7.7 & 4.2 & 2.5 \\
Targets 2014--2019 & 25 & 14.7 & 12.8 & 11.2 & 8.4 & 5.0 & 3.4 \\
26 industries, rebalanced & 25 & 11.6 & 10.0 & 8.6 & 6.1 & 3.2 & 1.9 \\

\bottomrule\end{tabular}
\tabnote{Premium of the own five-year mean over the five nearest peers, in points. The baseline is Table~\ref{tab:horizon}.}
\end{table}

\begin{table}[!htbp]\centering\small
\caption{Losses after fitting each forecast kernel to the recorded totals}\label{tab:ras}
\begin{tabular}{rrrrrlr}\toprule
 & \multicolumn{3}{c}{Kernel taken from} & \multicolumn{3}{c}{Template minus latest own table}\\
\cmidrule(lr){2-4}\cmidrule(lr){5-7}
Years ahead & Latest own table & Five peers & Template & Mean & 95\% interval & Positive \\
\midrule
1 & 4.6 & 19.3 & 19.6 & 15.0 & [13.3, 16.7] & 25/25 \\
2 & 6.6 & 19.3 & 19.6 & 13.0 & [11.3, 14.6] & 25/25 \\
3 & 8.1 & 19.3 & 19.6 & 11.5 & [9.7, 13.4] & 25/25 \\
5 & 10.6 & 19.5 & 19.6 & 9.0 & [7.0, 11.0] & 24/25 \\
8 & 14.1 & 19.8 & 19.6 & 5.5 & [3.0, 8.0] & 24/25 \\
10 & 15.7 & 20.2 & 19.6 & 3.9 & [1.6, 6.1] & 19/25 \\

\bottomrule\end{tabular}
\tabnote{\NnPanel{} economies, target years 2014 to 2020. Each kernel is fitted by RAS to the recorded row and column totals of the target year on the 34 industries. The loss is the total variation distance, in percent, between the fitted matrix and the recorded transaction shares. This is the error under the target year's uneven margins. It is not the kernel distance, which equals the RAS error only when the target totals are uniform.}
\end{table}

\begin{table}[!htbp]\centering\small
\caption{Own-history premium by buying industry}\label{tab:buyers}
\begin{tabular}{lrrrr}\toprule
 & & \multicolumn{3}{c}{Years ahead}\\
\cmidrule(lr){3-5}
Buying industries & Industries & 1 & 5 & 10 \\
\midrule
Manufacturing & 11 & 3.5 & 1.9 & 0.7 \\
Transport and storage & 5 & 2.8 & 1.6 & 0.7 \\
Public, education and health & 3 & 1.6 & 1.1 & 0.6 \\
Information and communication & 3 & 1.2 & 0.5 & 0.1 \\
Utilities & 2 & 1.1 & 0.6 & 0.3 \\
Finance and property & 2 & 1.0 & 0.5 & 0.2 \\
Other services & 2 & 0.9 & 0.5 & 0.2 \\
Professional and support services & 2 & 0.9 & 0.5 & 0.1 \\
Trade and hospitality & 2 & 0.8 & 0.5 & 0.2 \\
Agriculture and extraction & 1 & 0.4 & 0.3 & 0.2 \\
Construction & 1 & 0.4 & 0.2 & 0.1 \\
\midrule
All industries & 34 & 14.5 & 8.3 & 3.4 \\

\bottomrule\end{tabular}
\tabnote{\NnPanel{} economies, target years 2014 to 2020. Each cell of the kernel contributes half the absolute error of the peer forecast minus half the absolute error of the own five-year mean. Contributions are summed over suppliers and over the buying industries in each group, in points.}
\end{table}

\paragraph{Transitory and persistent change.} Suppose the observed kernel is a persistent component with uncorrelated increments of variance $s^2$ per year plus transitory noise of variance $w^2$. Then the expected squared Frobenius distance between tables $h$ years apart is $2w^2+s^2h$. A linear fit over ages one to eight in the measured panel gives a transitory share $2w^2/(2w^2+s^2)$ of \NnoiseShare{} percent of the one-year squared change. The estimate is descriptive: at ages above ten the squared distance grows more slowly than linearly.

\section{Two Databases}\label{app:databases}
Table~\ref{tab:matched} gives the numbers behind Figure~\ref{fig:databases}, adds the panel without the three economies that are template-like in some year, and reports forecasts scored on the other database. Table~\ref{tab:figaro} relates changes in FIGARO kernels to the documented origin of the tables.

\begin{table}[!htbp]\centering\small
\caption{The forecasting experiment in two databases}\label{tab:matched}
\setlength{\tabcolsep}{3.8pt}
\begin{tabular}{rrrrrrrrrrr}\toprule
 & \multicolumn{4}{c}{OECD} & \multicolumn{4}{c}{FIGARO} & \multicolumn{2}{c}{Premium, other database}\\
\cmidrule(lr){2-5}\cmidrule(lr){6-9}\cmidrule(lr){10-11}
Years & Latest & 5-year & Peers & Prem. & Latest & 5-year & Peers & Prem. & OECD history & FIGARO history \\
\midrule
\multicolumn{11}{l}{A. \NnBoth{} economies with national tables in both, targets 2015 to 2020}\\

1 & 6.8 & 9.2 & 22.2 & 13.0 & 5.5 & 8.4 & 24.6 & 16.2 & 9.4 & 7.2 \\
2 & 9.0 & 10.7 & 22.3 & 11.6 & 7.8 & 9.8 & 24.7 & 14.9 & 8.9 & 6.6 \\
3 & 10.5 & 12.0 & 22.5 & 10.5 & 9.3 & 11.1 & 24.7 & 13.6 & 8.3 & 6.0 \\
5 & 13.0 & 14.2 & 22.7 & 8.5 & 12.9 & 13.7 & 24.8 & 11.1 & 7.1 & 5.0 \\

\midrule
\multicolumn{11}{l}{B. \NnBothStrict{} of these whose OECD kernel is never template-like, targets 2015 to 2020}\\

1 & 6.3 & 8.8 & 22.3 & 13.5 & 5.6 & 8.6 & 24.2 & 15.6 & 9.3 & 8.0 \\
2 & 8.4 & 10.3 & 22.4 & 12.1 & 7.9 & 10.0 & 24.3 & 14.2 & 8.7 & 7.3 \\
3 & 10.0 & 11.6 & 22.5 & 11.0 & 9.4 & 11.4 & 24.4 & 12.9 & 8.1 & 6.7 \\
5 & 12.7 & 13.8 & 22.8 & 9.0 & 13.3 & 14.2 & 24.4 & 10.2 & 6.7 & 5.6 \\

\midrule
\multicolumn{11}{l}{C. All 42 economies, targets 2019 to 2022}\\

1 & 10.0 & 13.5 & 18.5 & 5.0 & 5.0 & 7.3 & 23.6 & 16.2 & 7.0 & \ensuremath{-}0.2 \\
2 & 14.5 & 15.1 & 18.8 & 3.6 & 7.0 & 8.8 & 23.6 & 14.9 & 6.6 & \ensuremath{-}0.4 \\
3 & 16.2 & 16.0 & 18.8 & 2.8 & 8.4 & 10.0 & 23.7 & 13.7 & 6.1 & \ensuremath{-}0.7 \\
5 & 17.2 & 17.3 & 18.9 & 1.6 & 11.0 & 12.0 & 23.8 & 11.8 & 5.0 & \ensuremath{-}1.1 \\

\bottomrule\end{tabular}
\tabnote{33 industries, histories from 2010. Losses in percent and premium in points. In the last two columns the histories of an economy and of its peers come from one database and the target kernel from the other.}
\end{table}

\begin{table}[!htbp]\centering\small
\caption{Change in FIGARO kernels by origin of the tables}\label{tab:figaro}
\begin{tabular}{lrrrr}\toprule
 & & \multicolumn{2}{c}{FIGARO} & OECD \\
\cmidrule(lr){3-4}\cmidrule(lr){5-5}
Origin of the FIGARO tables in the two years & Country-years & Mean & Median & Mean \\
\midrule
Delivered by national institutes in both years & 324 & 6.0 & 4.3 & 7.9 \\
Estimated by Eurostat in both years & 41 & 1.5 & 0.5 & 7.1 \\
Other combinations & 139 & 2.8 & 1.7 & 10.2 \\

\bottomrule\end{tabular}
\tabnote{42 economies, changes from the previous year for 2011 to 2022, total variation in percent. Origins are from \citet{eurostatmeta}. Of the 546 FIGARO country-years, \NnFigaroEstimated{} are Eurostat estimates. The last column gives the change in the OECD kernel for the same economies and years.}
\end{table}

\section{Multipliers}\label{app:multipliers}
Domestic input coefficients divide the domestic intermediate purchases of each of the 33 industry groups by the gross output of the buying group, using transactions and output from the same source and year. The OECD tables report gross output. For FIGARO we obtain it as the sum of each industry's deliveries to all intermediate and final users and check it against the sum of intermediate inputs and value added. All 1,092 systems have spectral radius below one. Multipliers are conditional on the retained domestic industries. They exclude imports and paths through excluded industries.

Table~\ref{tab:multyears} gives the annual comparison. In levels, a regression of the rank correlation on an indicator for a template-like OECD kernel with economy and year effects gives \NfeRhoClean{} [\NfeRhoCleanLow, \NfeRhoCleanHigh] for the \NnCleanMult{} economies that are specific in every year through 2019, and \NfeRhoAll{} [\NfeRhoAllLow, \NfeRhoAllHigh] for all 42. The eleven economies that are template-like in some earlier year are mostly ones whose FIGARO tables are themselves estimated or taken from the OECD, and their agreement between databases does not move with the template. This is the reason for comparing changes in the year of the switch in Table~\ref{tab:events}.

\begin{table}[!htbp]\centering\small
\caption{Agreement between OECD and FIGARO by year}\label{tab:multyears}
\setlength{\tabcolsep}{4pt}
\begin{tabular}{rrrrrrrrrr}\toprule
 & Template- & \multicolumn{2}{c}{Mean multiplier} & Absolute & \multicolumn{2}{c}{Rank correlation} & Common & Supplier & Kernel \\
\cmidrule(lr){3-4}\cmidrule(lr){6-7}
Year & like & OECD & FIGARO & gap & All 42 & \NnCleanMult{} & top five & shares & distance \\
\midrule
2010 & 5 & 1.593 & 1.581 & 0.087 & 0.794 & 0.801 & 3.62 & 18.2 & 16.8 \\
2011 & 3 & 1.588 & 1.571 & 0.088 & 0.779 & 0.788 & 3.55 & 18.0 & 16.9 \\
2012 & 4 & 1.584 & 1.569 & 0.085 & 0.790 & 0.798 & 3.33 & 19.2 & 18.3 \\
2013 & 4 & 1.584 & 1.565 & 0.081 & 0.804 & 0.807 & 3.43 & 18.0 & 16.9 \\
2014 & 4 & 1.582 & 1.564 & 0.082 & 0.794 & 0.801 & 3.36 & 17.8 & 16.5 \\
2015 & 5 & 1.577 & 1.564 & 0.079 & 0.807 & 0.805 & 3.45 & 17.3 & 16.0 \\
2016 & 5 & 1.575 & 1.562 & 0.082 & 0.798 & 0.792 & 3.43 & 18.3 & 17.1 \\
2017 & 5 & 1.570 & 1.559 & 0.084 & 0.800 & 0.798 & 3.50 & 18.6 & 16.9 \\
2018 & 2 & 1.567 & 1.553 & 0.084 & 0.785 & 0.786 & 3.48 & 18.9 & 17.4 \\
2019 & 5 & 1.567 & 1.552 & 0.082 & 0.792 & 0.790 & 3.52 & 19.2 & 17.5 \\
2020 & 8 & 1.567 & 1.557 & 0.085 & 0.789 & 0.796 & 3.48 & 19.0 & 17.5 \\
2021 & 25 & 1.569 & 1.558 & 0.101 & 0.743 & 0.736 & 3.26 & 23.3 & 22.6 \\
2022 & 38 & 1.564 & 1.563 & 0.107 & 0.725 & 0.708 & 3.26 & 25.8 & 25.4 \\

\bottomrule\end{tabular}
\tabnote{42 economies and 33 industries. Template-like counts OECD kernels. The absolute gap is the mean absolute difference between the two databases' multipliers for the same economy and industry. Rank correlations are within-economy Spearman correlations, averaged over economies. Common top five is the number of industries among the five largest multipliers in both databases. Supplier shares and kernel distance are total variation distances between the databases in percent.}
\end{table}

\clearpage
\bibliographystyle{plainnat}
{\small\setlength{\bibsep}{3pt}\bibliography{references}}

@techreport{carvalhovoigt,
 author={Carvalho, Vasco M. and Voigtl{\"a}nder, Nico},
 title={Input Diffusion and the Evolution of Production Networks},
 institution={National Bureau of Economic Research},
 type={Working Paper}, number={20025}, year={2014},
 note={Revised March 2015}, doi={10.3386/w20025}
}

@article{oberfield,
 author={Oberfield, Ezra},title={A Theory of Input--Output Architecture},
 journal={Econometrica},year={2018},volume={86},number={2},pages={559--589},doi={10.3982/ECTA10731}
}

@article{acemogluazar,
 author={Acemoglu, Daron and Azar, Pablo D.},title={Endogenous Production Networks},
 journal={Econometrica},year={2020},volume={88},number={1},pages={33--82},doi={10.3982/ECTA15899}
}

@article{mundt,
 author={Mundt, Philipp},title={The Formation of Input--Output Architecture: Evidence from the European Union},
 journal={Journal of Economic Behavior \& Organization},year={2021},volume={183},pages={89--104},
 doi={10.1016/j.jebo.2020.12.031}
}

@article{barros2026,
 author={Barros, Fernando and Delalibera, Bruno R. and Ribeiro, Marcos J. and Teignier, Marc},
 title={Production Networks and Structural Transformation},journal={Economic Modelling},year={2026},
 volume={156},pages={107461},doi={10.1016/j.econmod.2025.107461}
}

@article{reisch2026,
 author={Reisch, Tobias and Borsos, Andr{\'a}s and Thurner, Stefan},
 title={Supply Chain Network Rewiring Dynamics at the Firm Level}, journal={PNAS Nexus},
 year={2026},volume={5},number={4},pages={pgag091},doi={10.1093/pnasnexus/pgag091}
}

@article{nordlund,
 author={Nordlund, Carl},title={Transformations, Trajectories, and Similarities of National Production Structures: A Comparative Fingerprinting Approach},
 journal={PLOS ONE},year={2023},volume={18},number={12},pages={e0295568},doi={10.1371/journal.pone.0295568}
}

@article{haddad,
 author={Haddad, Eduardo A. and Ara{\'u}jo, In{\'a}cio F. and Perobelli, Fernando S.},
 title={Geographical Proximity and Technological Similarity},
 journal={Structural Change and Economic Dynamics},year={2024},volume={70},pages={302--318},doi={10.1016/j.strueco.2024.02.007}
}

@article{tarne,
 author={Tarne, Peter and Lehmann, Annekatrin and Finkbeiner, Matthias},
 title={A Comparison of Multi-Regional Input--Output Databases Regarding Transaction Structure and Supply Chain Analysis},
 journal={Journal of Cleaner Production},year={2018},volume={196},pages={1486--1500},doi={10.1016/j.jclepro.2018.06.082}
}

@article{pinero,
 author={Pi{\~n}ero, Pablo and Kutlina-Dimitrova, Zornitsa and Rueda-Cantuche, Jos{\'e} Manuel},
 title={{EU} Trade: Deviations in Multi-Country Input--Output Tables and Their Implications for Trade Policy},
 journal={National Accounting Review},year={2025},volume={7},number={4},pages={546--573},doi={10.3934/NAR.2025023}
}

@article{diem,
 author={Diem, Christian and Borsos, Andr{\'a}s and Reisch, Tobias and Kert{\'e}sz, J{\'a}nos and Thurner, Stefan},
 title={Estimating the Loss of Economic Predictability from Aggregating Firm-Level Production Networks},
 journal={PNAS Nexus},year={2024},volume={3},number={3},pages={pgae064},doi={10.1093/pnasnexus/pgae064}
}

@article{sinkhorn,
 author={Sinkhorn, Richard and Knopp, Paul},title={Concerning Nonnegative Matrices and Doubly Stochastic Matrices},
 journal={Pacific Journal of Mathematics},year={1967},volume={21},number={2},pages={343--348},doi={10.2140/pjm.1967.21.343}
}

@article{chakrabarty,
 author={Chakrabarty, Deeparnab and Khanna, Sanjeev},title={Better and Simpler Error Analysis of the {Sinkhorn--Knopp} Algorithm for Matrix Scaling},
 journal={Mathematical Programming},year={2021},volume={188},pages={395--407},doi={10.1007/s10107-020-01503-3}
}

@article{cameron,
 author={Cameron, A. Colin and Gelbach, Jonah B. and Miller, Douglas L.},title={Robust Inference with Multiway Clustering},
 journal={Journal of Business \& Economic Statistics},year={2011},volume={29},number={2},pages={238--249},doi={10.1198/jbes.2010.07136}
}

@article{mackinnon,
 author={MacKinnon, James G. and Nielsen, Morten {\O}rregaard and Webb, Matthew D.},
 title={Cluster-Robust Inference: A Guide to Empirical Practice},journal={Journal of Econometrics},
 year={2023},volume={232},number={2},pages={272--299},doi={10.1016/j.jeconom.2022.04.001}
}

@misc{oecd2025,
 author={{OECD}},title={Inter-Country Input--Output Tables, Regular 2025 Edition},
 year={2025},url={https://www.oecd.org/en/data/datasets/inter-country-input-output-tables.html}
}

@techreport{oecd2026,
 author={{OECD}},title={Trends in Global Value Chains},
 institution={OECD Publishing},address={Paris},year={2026},doi={10.1787/8217c2ff-en}
}

@misc{eurostat,
 author={{Eurostat}},title={{FIGARO}: {EU} Inter-Country Supply, Use and Input--Output Tables, 2026 Edition},
 year={2026},url={https://ec.europa.eu/eurostat/en/web/esa-supply-use-input-tables/database}
}

@article{tashman,
 author={Tashman, Leonard J.},
 title={Out-of-Sample Tests of Forecasting Accuracy: An Analysis and Review},
 year={2000},journal={International Journal of Forecasting},volume={16},number={4},pages={437--450},
 doi={10.1016/S0169-2070(00)00065-0}
}

@techreport{yamano,
 author={Yamano, Norihiko and Alsamawi, Ali and Webb, Colin and Cimper, Agnès and Zürcher, Carmen and Chiapin Pechansky, Ricardo},
 title={Development of the {OECD} Inter Country Input--Output Database 2023},
 institution={OECD Publishing},series={OECD Science, Technology and Industry Working Papers},number={2023/08},
 year={2023},doi={10.1787/5a5d0665-en}
}

@book{unhandbook,
 author={{United Nations}},
 title={Handbook on Supply, Use and Input--Output Tables with Extensions and Applications},
 year={2018},publisher={United Nations},
 url={https://unstats.un.org/unsd/nationalaccount/docs/SUT_IOT_HB_Final_Cover.pdf}
}

@incollection{timmermann2006,
  author = {Timmermann, Allan}, title = {Forecast Combinations}, year = {2006},
  booktitle = {Handbook of Economic Forecasting}, editor = {Elliott, Graham and Granger, Clive W. J. and Timmermann, Allan},
  publisher = {Elsevier}, volume = {1}, pages = {135--196}, doi = {10.1016/S1574-0706(05)01004-9}
}

@article{pesaran2026,
  author = {Pesaran, M. Hashem and Pick, Andreas and Timmermann, Allan}, year = {2026},
  title = {Forecasting with panel data: Estimation uncertainty versus parameter heterogeneity},
  journal = {Quantitative Economics}, volume = {17}, number = {2}, pages = {342--393}, doi = {10.3982/QE2589}
}

@article{beaufils2022,
  author = {Beaufils, Timoth{\'e} and Wenz, Leonie}, year = {2022},
  title = {A scenario-based method for projecting multi-regional input-output tables},
  journal = {Economic Systems Research}, volume = {34}, number = {4}, pages = {440--468}, doi = {10.1080/09535314.2021.1952404}
}

@article{huo2022,
  author = {Huo, Jingwen and Chen, Peipei and Hubacek, Klaus and Zheng, Heran and Meng, Jing and Guan, Dabo}, year = {2022},
  title = {Full-scale, near real-time multi-regional input-output table for the global emerging economies ({EMERGING})},
  journal = {Journal of Industrial Ecology}, volume = {26}, number = {4}, pages = {1218--1232}, doi = {10.1111/jiec.13264}
}

@article{owen2014,
  author = {Owen, Anne and Steen-Olsen, Kjartan and Barrett, John and Wiedmann, Thomas and Lenzen, Manfred}, year = {2014},
  title = {A structural decomposition approach to comparing {MRIO} databases},
  journal = {Economic Systems Research}, volume = {26}, number = {3}, pages = {262--283}, doi = {10.1080/09535314.2014.935299}
}

@article{badr2026,
  author = {Badr, Mohamed and Stadler, Konstantin and Ottelin, Juudit}, year = {2026},
  title = {A comprehensive review of uncertainty in multi-regional input-output modelling},
  journal = {Ecological Modelling}, volume = {519}, pages = {111649}, doi = {10.1016/j.ecolmodel.2026.111649}
}

@article{lenzen2011,
  author = {Lenzen, Manfred}, year = {2011},
  title = {Aggregation Versus Disaggregation in Input-Output Analysis of the Environment},
  journal = {Economic Systems Research}, volume = {23}, number = {1}, pages = {73--89},
  doi = {10.1080/09535314.2010.548793}
}

@article{acemoglu2012,
  author = {Acemoglu, Daron and Carvalho, Vasco M. and Ozdaglar, Asuman and Tahbaz-Salehi, Alireza},
  title = {The Network Origins of Aggregate Fluctuations}, journal = {Econometrica}, year = {2012},
  volume = {80}, number = {5}, pages = {1977--2016}, doi = {10.3982/ECTA9623}
}

@article{atalay,
  author = {Atalay, Enghin}, title = {How Important Are Sectoral Shocks?},
  journal = {American Economic Journal: Macroeconomics}, year = {2017}, volume = {9}, number = {4},
  pages = {254--280}, doi = {10.1257/mac.20160353}
}

@article{baqaeefarhi,
  author = {Baqaee, David Rezza and Farhi, Emmanuel},
  title = {The Macroeconomic Impact of Microeconomic Shocks: Beyond {Hulten's} Theorem},
  journal = {Econometrica}, year = {2019}, volume = {87}, number = {4}, pages = {1155--1203}, doi = {10.3982/ECTA15202}
}

@article{caliendoparro,
  author = {Caliendo, Lorenzo and Parro, Fernando}, title = {Estimates of the Trade and Welfare Effects of {NAFTA}},
  journal = {Review of Economic Studies}, year = {2015}, volume = {82}, number = {1}, pages = {1--44},
  doi = {10.1093/restud/rdu035}
}

@article{johnson2018,
  author = {Johnson, Robert C.}, title = {Measuring Global Value Chains}, journal = {Annual Review of Economics},
  year = {2018}, volume = {10}, pages = {207--236}, doi = {10.1146/annurev-economics-080217-053600}
}

@article{chenery1958,
  author = {Chenery, Hollis B. and Watanabe, Tsunehiko}, title = {International Comparisons of the Structure of Production},
  journal = {Econometrica}, year = {1958}, volume = {26}, number = {4}, pages = {487--521}, doi = {10.2307/1907514}
}

@book{stone1961,
  author = {Stone, Richard}, title = {Input--Output and National Accounts}, year = {1961},
  publisher = {Organisation for European Economic Co-operation}, address = {Paris}
}

@book{bacharach1970,
  author = {Bacharach, Michael}, title = {Biproportional Matrices and Input--Output Change}, year = {1970},
  publisher = {Cambridge University Press}, address = {Cambridge}
}

@article{lahr2004,
  author = {Lahr, Michael L. and de Mesnard, Louis},
  title = {Biproportional Techniques in Input--Output Analysis: Table Updating and Structural Analysis},
  journal = {Economic Systems Research}, year = {2004}, volume = {16}, number = {2}, pages = {115--134},
  doi = {10.1080/0953531042000219259}
}

@article{temurshoev2011,
  author = {Temurshoev, Umed and Webb, Colin and Yamano, Norihiko},
  title = {Projection of Supply and Use Tables: Methods and Their Empirical Assessment},
  journal = {Economic Systems Research}, year = {2011}, volume = {23}, number = {1}, pages = {91--123},
  doi = {10.1080/09535314.2010.534978}
}

@article{wiebe2016,
  author = {Wiebe, Kirsten S. and Lenzen, Manfred},
  title = {To {RAS} or Not to {RAS}? {What} Is the Difference in Outcomes in Multi-Regional Input--Output Models?},
  journal = {Economic Systems Research}, year = {2016}, volume = {28}, number = {3}, pages = {383--402},
  doi = {10.1080/09535314.2016.1192528}
}

@article{johnson2013,
  author = {Johnson, Simon and Larson, William and Papageorgiou, Chris and Subramanian, Arvind},
  title = {Is Newer Better? {Penn World Table} Revisions and Their Impact on Growth Estimates},
  journal = {Journal of Monetary Economics}, year = {2013}, volume = {60}, number = {2}, pages = {255--274},
  doi = {10.1016/j.jmoneco.2012.10.022}
}

@article{headries,
  author = {Head, Keith and Ries, John},
  title = {Increasing Returns versus National Product Differentiation as an Explanation for the Pattern of {U.S.}--{Canada} Trade},
  journal = {American Economic Review}, year = {2001}, volume = {91}, number = {4}, pages = {858--876},
  doi = {10.1257/aer.91.4.858}
}

@article{mosteller,
  author = {Mosteller, Frederick}, title = {Association and Estimation in Contingency Tables},
  journal = {Journal of the American Statistical Association}, year = {1968}, volume = {63}, number = {321},
  pages = {1--28}, doi = {10.1080/01621459.1968.11009219}
}

@techreport{oecdnotes2025,
  author = {Yamano, Norihiko and Webb, Colin and Cimper, Agn{\`e}s and Pazos, Rodrigo and Kohli, Akash},
  title = {Development of {OECD} Inter-Country Input--Output Database, 2025 Edition: Supplementary Notes},
  institution = {OECD Directorate for Science, Technology and Innovation}, year = {2025}, type = {Technical note},
  url = {https://www.oecd.org/en/data/datasets/inter-country-input-output-tables.html}
}

@misc{eurostatmeta,
  author = {{Eurostat}},
  title = {{EU} Inter-Country Supply, Use and Input--Output Tables ({FIGARO}), Current Prices (naio\_10\_fcp): Reference Metadata},
  year = {2026}, url = {https://ec.europa.eu/eurostat/cache/metadata/en/naio_10_fcp_esms.htm}
}
\end{document}